\documentclass[reprint,
groupedaddress,
amsmath,amssymb,
aps,
prb,
final,
nolongbibliography,
]{revtex4-2}

\usepackage[T1]{fontenc}
\usepackage{amsmath}
\usepackage{graphicx}
\usepackage{bm,upgreek}
\usepackage{xcolor}
\definecolor{mred}{HTML}{CD2B44}
\definecolor{mblue}{HTML}{2C44CE}
\usepackage[
colorlinks,
citecolor=mred,
linkcolor=mblue,
urlcolor=black,
]{hyperref}

\usepackage{dsfont}

\begin{document}
         \title{Quantized charge transport in disordered Floquet topological
   insulators in the absence of Anderson localization}
	\author{Lennard Berg}\email{lennard.berg98@gmail.de}
	\affiliation{Institute of Physics, University of Greifswald, Felix-Hausdorff-Straße 6, 17489 Greifswald, Germany}	
	\author{Andreas Alvermann}\email{alvermann@physik.uni-greifswald.de} \affiliation{Institute of Physics, University of Greifswald, Felix-Hausdorff-Straße 6, 17489 Greifswald, Germany}	
	\author{Holger Fehske}\email{fehske@physik.uni-greifswald.de}
	\affiliation{Institute of Physics, University of Greifswald, Felix-Hausdorff-Straße 6, 17489 Greifswald, Germany}
	\affiliation{Erlangen National High Performance Computing Center, 91058 Erlangen, Germany}

	\date{\today}
	
	\begin{abstract}
		We perform a numerical study of Floquet topological insulators with temporal disorder to investigate the existence of quantized charge transport without Anderson localization. We first argue that in setups with temporal imperfections Anderson localization can not be expected but bulk transport is diffusive in the long-time limit. 
		In a second step we compute the corrections to the cumulative averaged pumped charge due to the temporal disorder and show that transport is characterized by two regimes: the transient regime, represented by a plateau for uncorrelated disorder, and the long-time behavior with a common scaling law for both uncorrelated and correlated disorder. 
		Most notably, our numerical results indicate that the dynamic corrections vanish in the long-time limit such that quantized charge transport and diffusive bulk motion can coexist in temporally disordered Floquet topological insulators. 
	\end{abstract}
	
	\maketitle
	
	\section{\label{sec:intro}Introduction}
	The relevance of topological properties for charge transport became initially clear with the quantum Hall effect \cite{RevModPhys.58.519}. In the quantum Hall effect, topologically protected chiral edge states give rise to lossless charge transport which is quantized and linked to a topological index \cite{Thouless,TKNN}. 
Novel topological states of matter result from symmetries, non-Hermiticiy, or periodic driving~\cite{%
PhysRevB.82.235114,%
HasanKane2010,%
RevModPhys.83.1057_11,%
PhysRevX.3.031005,%
nathan2015topological,%
RevModPhys.88.035005_16,%
basov2017towards,%
zhang2017observation,%
PhysRevB.97.045140,%
PhysRevB.97.245401,%
oka2019floquet,%
PhysRevLett.123.190403,%
PhysRevB.99.195133,%
PhysRevB.99.235408,%
PhysRevB.99.245102,%
wintersperger2020realization,PhysRevX.10.021032,PhysRevB.101.041403,%
Fedorova2020,%
PhysRevResearch.2.023235,%
bergholtz2020exceptional%
}.
 In particular, so-called Floquet engineering gives active control over the topological properties of matter  \cite{dahlhaus2015magnetization,kaladzhyan2017controlling,PhysRevA.98.013635,topp2022orbital} as confirmed by recent experiments for photonic \cite{rechtsman2013photonic,maczewsky2017observation,mukherjee2017experimental,RevModPhys.91.015006_19,NatMat20}, acoustic \cite{fleury2016floquet,peng2016experimental} and electronic systems \cite{nagulu2022chip,kumar2022topological}, and can induce topological phases without a static counterpart.
In Floquet insulators, the inherent non-adiabatic effects of the periodic driving generally cause deviations from quantized charge transport \cite{Shih,PhysRevB.92.165111,Privitera}.
	
The so-called anomalous Floquet-Anderson insulator (AFAI), however, can exhibit quantized topological protected edge transport as well as quantized magnetization density \cite{Titum1,PhysRevB.101.165401,PhysRevResearch.2.022048,timms2021quantized}. The key point is that disorder leads to Anderson localization, such that in a topological phase localized bulk states coexist with delocalized edge states at the same quasi-energies.

	In experimental Floquet setups, temporal fluctuations are equally likely as static disorder. Because of temporal fluctuations, bulk states exhibit  diffusive behavior rather than strict Anderson localization~\cite{%
PhysRevLett.80.4111,%
PhysRevLett.81.1203,%
PhysRevA.67.042315,%
kendon_2007,%
yin2008quantum,%
schreiber2011decoherence,%
white2014phase,vcadevz2017dynamical,PhysRevB.98.214301,PhysRevLett.120.216801,%
ravindranath2021dynamical,cao2021interaction}.
For such a situation, the only available study known to us reveals the existence of plateaus for the pumped charge~\cite{timms2021quantized}.
However, the plateau values reported in this reference are not quantized because a simpler ``one-dimensional'' disorder, which still preserves translational symmetry along the edge of the system, is considered. This leaves open the question whether quantized plateaus occur for full ``two-dimensional'' disorder, which breaks translational symmetry in all directions.

Note that ``two-dimensional'' disorder is required for Anderson localization, that is for quantization of the pumped charge without temporal fluctuations.
Therefore, the question whether quantization survives temporal fluctuations or is lost together with Anderson localization, makes sense only if one studies a fully disordered system. In the present work, we perform this study.

	Our investigation is based on the numerical analysis of transport in a disordered Floquet topological insulator (DFTI) with uncorrelated or correlated temporal disorder. We first reexamine the effects of temporal disorder to establish that transport is indeed diffusive in the DFTIs under consideration~\cite{vcadevz2017dynamical,PhysRevB.98.214301,PhysRevLett.120.216801,ravindranath2021dynamical}. Then, we compute the transported charge per period for a sequence of arbitrary propagators, adapting the approach from Ref.~\cite{Titum1}.  	
In agreement with the results from the literature~\cite{timms2021quantized} we observe non-quantized plateaus, where the pumped charge deviates from one. 
But for the ``two-dimensional'' disorder used here these plateaus have only finite lifetime; moving beyond the transient regime we observe a transition to a characteristic long-time behavior described by a common scaling law indicative of quantization.
The scaling law holds on time scales restricted only by the finite size of the systems studied numerically.
 Extrapolation of the scaling law to infinite systems gives a strong argument for the emergence of strictly quantized charge transport in the long-time limit. For uncorrelated temporal disorder, we can even provide a phenomenological ansatz to describe the deviations from quantized transport in the transition regime.

	\section{\label{sec:model}Model}
\begin{figure}
		\centering
		\includegraphics{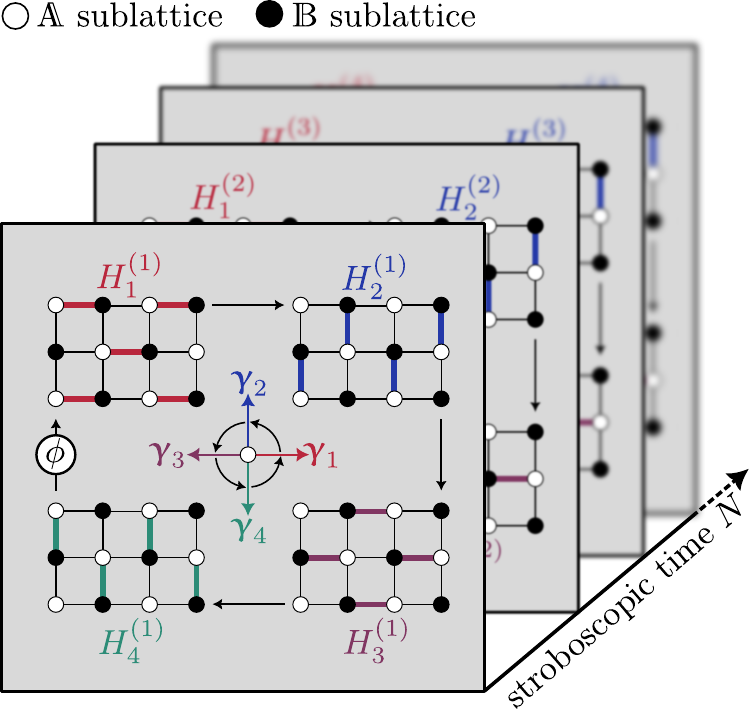}
		\caption{\label{fig:model} Two-dimensional tight-binding model of a DFTI with additional temporal disorder. In each period (each gray layer), the Hamiltonian is piecewise constant on a segment with duration \(T/4\). The nearest-neighbor hopping in each segment is represented by the corresponding direction \(\bm{\upgamma}_i\) (colored bonds) and the coupling constant \(\mathcal{J}^{(n)}\). The latter changes from period to period according to  Eq.~\eqref{eq:JDist}. At the end of each period, a \(\delta\)-kick partially randomizes the phase.}
	\end{figure}

	From now on, we consider a DFTI with (uncorrelated or correlated) temporal disorder on a bipartite square lattice \(\Gamma\)  with \(L_x\times L_y\) sites. The time-periodic driving is encoded by a 4-step protocol, see Fig.~\ref{fig:model}. In each step of the protocol, only pairwise coupling of lattice sites (between the \(\mathds{A}\) and \(\mathds{B}\) sublattice) is allowed. The directions of the coupling are defined as \((\bm{\upgamma}_1,\bm{\upgamma}_2,\bm{\upgamma}_3,\bm{\upgamma}_4)=(\bm{e}_1,-\bm{e}_2,-\bm{e}_1,\bm{e}_2)\), where \(\bm{e}_{1,2}\) are the standard basis vectors of \(\mathds{R}^2\) \cite{PhysRevX.3.031005}. Temporal disorder is introduced by varying the coupling constants \(\mathcal{J}=\mathcal{J}^{(n)}\) in each period. 
	
	The Hamiltonian of the \(n-\)th period reads  
	\begin{equation}
		H^{(n)}(t)=H^{(n)}_k+\delta(t-nT)H_{\mathrm{stat-dis}}
	\end{equation}
	for \((k-1)\leq4[t/T-(n-1)]<k\) (\(k\in\{1,2,3,4\}\)) on four segments of duration $T/4$, with 
	\begin{equation}
		H^{(n)}_k=\mathcal{J}^{(n)}\sum_{\mathbf{r}\in \Gamma}(c_{\mathbf{r}+\bm{\upgamma}_k}^\dagger c_{\mathbf{r}}+ c_{\mathbf{r}}^\dagger c_{\mathbf{r}+\bm{\upgamma}_k})\,.
	\end{equation}
Here, \(c_{\bm{r}}^\dagger\)  (\(c_{\bm{r}}^{}\)) is the fermionic creation (annihilation) operator on lattice site \(\mathbf{r}\). At the end of each period, a \(\delta\)-kick with 
\begin{equation}
	H_{\mathrm{stat-dis}}=\sum_{\mathbf{r}\in\Gamma}\phi_{\mathbf{r}}c_{\mathbf{r}}^\dagger c_{\mathbf{r}}
\end{equation}
acts as static disorder. We choose \((\phi_{\mathbf{r}})_{\mathbf{r}\in\Gamma}\) as uniformly distributed random variables in the interval \([-\pi,\pi]\). Without temporal disorder, this so-called phase disorder is suitable to achieve Anderson localization \cite{PhysRevB.101.165401,PhysRevResearch.2.022048,yin2008quantum,hamza2009dynamical}. Note that other types of static disorder, such as random on-site potentials, can also be used \cite{Titum1,timms2021quantized}. 
Importantly, the disorder considered here breaks translational symmetry in all directions (``two-dimensional'' disorder in the classification of Ref.~\cite{timms2021quantized}). Without temporal fluctuations, the disorder leads to Anderson localization. With temporal fluctuations, Anderson localization is lost, as will be discussed later in more detail.

	To study the dynamics generated by the sequence of Hamiltonians \((H^{(n)}(t))\), we consider  the corresponding sequence of propagators  
	\begin{eqnarray}
		U^{(n)}(T)&=&\mathcal{T}\exp\left\{-i\int_{(n-1)T}^{nT} H^{(n)}(t)\mathrm{d}t\right\}\nonumber\\&=&S_\phi U^{(n)}_4U^{(n)}_3U^{(n)}_2U^{(n)}_1\,,
	\end{eqnarray}
	for each period. Here, \(U_k^{(n)}=e^{-iTH^{(n)}_k/4}\), \(S_\phi=e^{-iH_{\mathrm{stat-dis}}}\), and \(\mathcal{T}\) is the time-ordering operator. The propagator over \(N\) periods is \(U(NT)=\prod_{n=1}^{N} U^{(n)}(T)\), where \(N T\) is the stroboscopic time for a given number of periods. In our numerical computations we set \(T=4\) and simplify the notation by dropping the factor \(T\) in the argument, thus measuring time in units of \(T\).
	
	The temporal variation of the coupling constant is supposed to model frequency fluctuations, e.g., of the driving laser in a hypothetical experiment. Within each period \(\mathcal{J}^{(n)}\) is obtained from a normal distribution
		\begin{equation}
			P(\mathcal{J}^{(n)}=\mathcal{J})= \frac{1}{\sqrt{2\pi\eta^2}}e^{-\frac{(\mathcal{J}-\mathcal{J}_0)^2}{2\eta^2}} \quad (\text{for } n>1)\,,
			\label{eq:JDist}
		\end{equation}
		where \(\eta\) parametrizes the disorder strength. Only for \(n=1\) we use a fixed value \(\mathcal{J}^{(1)}=\mathcal{J}_0\).
		
		 Without temporal disorder, the present model of a (D)FTI exhibits a phase transition in the thermodynamic limit at \(\mathcal{J}=\pi/4\). For values \(\mathcal{J}<\pi/4\), the transported charge is equal to zero, and for \(\mathcal{J}>\pi/4\), it is equal to one. To investigate the possibility of quantization in the presence of temporal disorder it is advantageous that each individual propagator \(U^{(n)}\)
	  leads to quantized transport if used for propagation over multiple periods. To simplify further, we will set \(\mathcal{J}_0 = \pi/2\), which corresponds to perfect coupling. In this specific case, \(U^{(1)}\) acts as identity operator on bulk states and as shift operator on edge states. This choice is arbitrary but representative. 
	
	To show that the following numerical results are not restricted to the specific type of temporal disorder chosen here, we present additional results for a different model in Appendix~\ref{appendix:missed}.
	
	\section{\label{sec:numerical}Numerical approach, results and discussion}
	\subsection{Absence of Anderson localization in the long-time limit}
	In this subsection, we study the behavior of bulk transport under the influence of temporal disorder. For quantum walks \cite{kendon_2007,PhysRevA.67.042315,yin2008quantum,schreiber2011decoherence}, quantum kicked rotors \cite{white2014phase,PhysRevLett.80.4111,PhysRevLett.81.1203,cao2021interaction} and also (D)FTIs \cite{ravindranath2021dynamical,vcadevz2017dynamical,PhysRevB.98.214301,PhysRevLett.120.216801} it is known that temporal disorder leads to diffusive transport in the bulk, i.e., to a breakdown of Anderson localization inspite of disorder. In the following,  we present results for a DFTI with temporal disorder to show that a characteristic scaling of the pumped charge appears together with the diffusive bulk transport and extends into the long-time limit.
	
	For a wave packet with variance \(\sigma^2(N)\), written as a function of stroboscopic time $N$, the asymptotic dynamics for \(N\to\infty\)  is described by the relation for anomalous diffusion \cite{metzler2014anomalous}
	\begin{equation}
		\sigma^2(N)=KN^{\beta}\,,
		\label{eq:AnoDiff}
	\end{equation}
	where \(K\) is the (generalized) diffusion constant. The anomalous diffusion exponent \(\beta\) defines different transport regimes: \(\beta=0\) (localized), \(\beta=1\) (diffusive) and \(\beta=2\) (ballistic). 
	
	To determine the bulk transport regime, we compute the variance \(\sigma^2(N)\) from the transmission probability
	\begin{equation}
		\mathcal{G}_{ji}(N)=|\langle j | U(N)|i\rangle |^2
		\label{eq:BulkTrans}
	\end{equation}
	 from an initial lattice site (\(|i\rangle=|(x_0,y_0)\rangle\))  to another lattice site $j$.
	 The variance \(\sigma_i^2(N)\) is  the second moment   
		\begin{equation}
			\sigma_i^2(N)=\sum_{j=(x,y)\in\Gamma}[(x-x_0)^2+(y-y_0)^2]\mathcal{G}_{ji}(N) \;.
		\end{equation}
		Instead of the variance, we can equally consider the spread \(\sigma_i(N)\).
		
		To remove the dependence of $\sigma_i(N)$ on the initial lattice site $i$, we average over initial sites, e.g., $\sigma(N)= (L_x L_y)^{-1} \sum_{i \in \Gamma} \sigma_i(N)$ for an average over the entire lattice.
	Without temporal disorder, \(\sigma(N)\simeq \mathrm{const.}\) for all \(N\) due to  Anderson localization. In general, \(\sigma(N)\) can fluctuate around a constant mean value. When adding temporal disorder, a transition emerges where the exponent $\beta$ in Eq.~\eqref{eq:AnoDiff} gradually changes from \(\beta=0\) to \(\beta=1\) around a certain time \(N_c\). Note that the transition time \(N_c\) is an approximate quantity, rather than the specific time of a sharp transition, and depends on the disorder strength. 
	
	The spread \(\sigma(N)\) can be used to monitor the topological breakdown appearing in the pumped charge. In the Anderson localized regime, finite-size effects are of the form \(e^{-L_{x,y}/\xi}\), where \(\xi\) denotes the (typical) localization length. In our case, with a transition towards diffusive transport, we can not define an appropriate localization length, but still use the quantity \(e^{-L_{x,y}/\sigma(N)}\)
to estimate the finite-size effects.
	
	In the numerical simulations, we take the four most central lattice sites as initial sites to minimize finite size effects. Adding more sites does not  change the results. The lattice used consists of $80\times 80$ sites. The sequence \((\mathcal{J}^{(n)})\) of random couplings is  generated from the probability distribution in Eq.~\eqref{eq:JDist}. We use two different values \(\mathcal{J}_0 \in \{ \pi/2, 0.9\times \pi/2\}\), to avoid restricting our discussion to the particular perfect-coupling value $\mathcal{J}_0 = \pi/2$. Finally, we average the spread \(\sigma(N)\) over $10^3$ disorder realizations.

\begin{figure}
		\centering
		\includegraphics{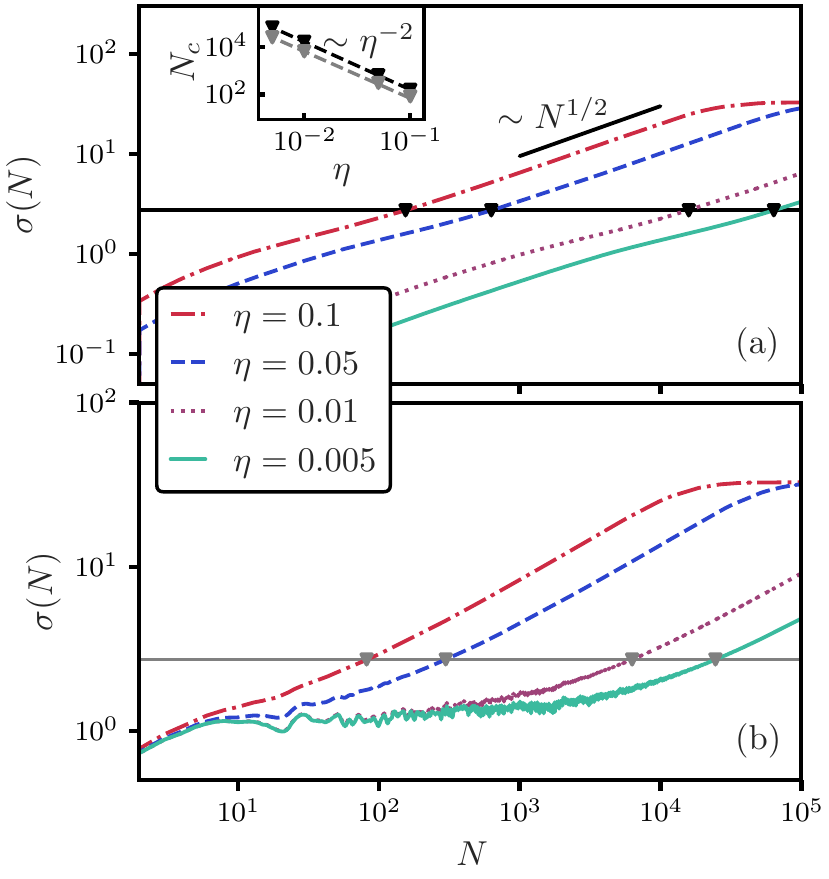}
		\caption{\label{fig:BulkSpread} Spread \(\sigma(N)\) as a function of stroboscopic time \(N\) at various disorder strengths \(\eta\). Results are given for two different mean values of the normal distribution~\eqref{eq:JDist}: (a) \(\mathcal{J}_0=\pi/2\) and (b) \(\mathcal{J}_0=0.9\times\pi/2\). The black and gray lines mark \(\sigma(N)=3\) in (a) and (b), respectively, which are used to determine \(N_c\). Inset: Scaling of \(N_c\) (defined by \(\sigma(N_c)=3\)) as a function of \(\eta\) [black and gray lines refer to panel (a) and (b), respectively.]}
	\end{figure}

For \(\mathcal{J}_0=\pi/2\), $\sigma(N)$ shows diffusive scaling $\sigma \sim N^{1/2}$
 for all times even if \(\sigma(N)\ll 1\), see  Fig.~\ref{fig:BulkSpread} (a). For $\sigma(N) \ll 1$ the state is ``quasi-localized'' at the initial lattice site and has almost no spread. We can safely interpret this situation in the way that diffusive behavior sets in as soon as the spread becomes comparable to the lattice constant, i.e., \(\sigma(N) \gtrsim 1\). In contrast, for \(\mathcal{J}_0=0.9\times \pi/2\) (and similarly for any \(\mathcal{J}_0\neq \pi/2\)), diffusive scaling is observed indeed only after the ``quasi-localized'' regime, see Fig.~\ref{fig:BulkSpread} (b). This allows us to identify the typical time $N_c$ of the transition from ``quasi-localized'' to diffusive transport.	In any case, the transition to diffusive scaling
  $\sigma \sim N^{1/2}$
after the transient regime is a direct consequence of temporal disorder.
	
	In order to obtain a concrete value for the typical time $N_c$, and afterwards of the dependence of $N_c$ on the disorder strength $\eta$ shown in the inset of Fig.~\ref{fig:BulkSpread}, we read off the point where 	\(\sigma(N_c)=3\). Of course, the qualitative dependence $N_c(\eta)$ does not depend on this particular choice.
	
	The curves in the inset of Fig.~\ref{fig:BulkSpread}  show  that the transition from quasi-localized to diffusive behavior occurs for any non-vanishing \(\eta\). The corresponding functional scaling is found to be \(N_c\sim \eta^{-2}\), and is the same for both values of \(\mathcal{J}_0\). The curve for \(\mathcal{J}_0=0.9\times \pi/2\) is only slightly shifted towards smaller values of $N_c$, i.e., to earlier times for the transition to diffusive transport.
	
	\subsection{Quantization of transported charge}
\begin{figure}
		\centering
		\includegraphics{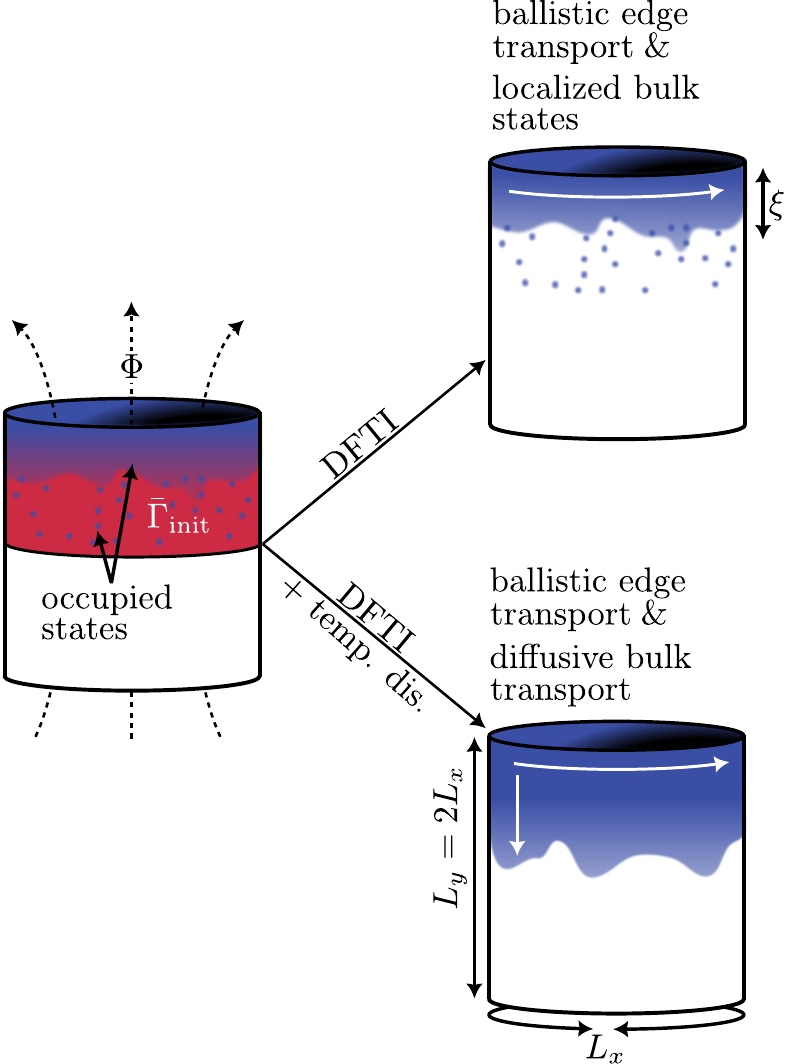}
		\caption{\label{fig:geometry} Left: Schematic representation of the geometry \(\bar{\Gamma}\) with threaded flux \(\Phi\) and initial occupied lattice region \(\bar{\Gamma}_{\mathrm{init}}\) (shaded red), which defines implicitly the initially occupied Floquet states (shaded blue). Right: Schematic representation of the time evolution of the initial states without temporal disorder (top) and with temporal disorder (bottom). Note that the flux was omitted for greater clarity on the right.} 
	\end{figure}

	We now prepare the system as depicted in Fig.~\ref{fig:geometry} (cf. Ref.~\cite{Titum1}).  We use a lattice geometry \(\bar{\Gamma}\) with periodic boundary conditions along the \(x\)-direction and consider a flux \(\Phi\) threaded through the cylinder. The flux is implemented by a phase factor \(e^{i\Phi}\) attached to the hopping matrix elements across the line \(x=x_0\). In other words, we use twisted boundary conditions (TBC)~\cite{Thouless2}. With the phase factor, the current operator across the line \(x_0\) can be written as \(j^{(n)}(t)=\partial_\Phi H^{(n)}_\Phi(t)\)~\cite{Titum1}, where \(H_\Phi^{(n)}\) denotes the Hamiltonian with TBC for the $n$-th period. The associated propagator is \(U_\Phi^{(n)}\).

	  The charge transported during the \(n\)-th period is given as expectation value of \(j^{(n)}(t)\) averaged over the respective period,
	\begin{equation}
		Q^{(n)}=\int_{(n-1)}^{n}\mathrm{Tr}\{\rho^{(n)}(t) j^{(n)}(t)\}\mathrm{d}t\,,
		\label{eq:ChargeBase}
	\end{equation}
	where \(\rho^{(n)}(t)\) is the corresponding density matrix. 
	
	The initial density matrix is
	\begin{equation}
		\rho^{(1)}(t)=\sum_{\alpha\alpha^\prime}c_{\alpha\alpha^\prime}|\psi_\alpha(t)\rangle\langle \psi_{\alpha^\prime}(t)|
	\end{equation}
	with
	\begin{equation}	
		c_{\alpha\alpha^\prime}=\sum_{\mathbf{r}\in\bar{\Gamma}_{\mathrm{init}}}\langle \psi_\alpha(0)|\mathbf{r}\rangle\langle \mathbf{r}|\psi_{\alpha^\prime}(0)\rangle\,,
	\end{equation}
	where \(|\psi_\alpha(t)\rangle\) is a Floquet state of \(U^{(1)}_\Phi(1)\) and \(\bar{\Gamma}_{\mathrm{init}}=\Bar{\Gamma}|_{y\geq L_y/2}\) represents the lattice 
	sites in the upper half of the cylinder [c.f., Fig.~\ref{fig:geometry} (a)]. The subsequent density matrices are obtained recursively,
		\begin{equation}
			\rho^{(n)}(t)=U_\Phi^{(n)}(t)\rho^{(n-1)}(1)U_\Phi^{(n)}(t)\,.
	\end{equation}
	Evaluation of the integral~\eqref{eq:ChargeBase} gives the charge
	\begin{equation}
		Q^{(n)}=i\mathrm{Tr}\{\rho^{(n)}(1) (U_\Phi^{(n)}(1))^\dagger \partial_\Phi U_\Phi^{(n)}(1)\}
		\label{eq:ChargeTempBase}
	\end{equation} 
	transported during the \(n\)-th period.
	Here, the trace is taken over all states \(|\psi_\alpha^{(n)}\rangle=U_\Phi^{(n)}|\psi_\alpha^{(n-1)}\rangle\) with \(|\psi_\alpha^{(1)}\rangle = |\psi_\alpha(0)\rangle\). Equation~\eqref{eq:ChargeTempBase} contains two parts: \((U_\Phi^{(n)}(1))^\dagger\partial_\Phi U^{(n)}_\Phi(1)\), a function of \(\Phi\), is linked to the winding number of the quasi-energy spectrum \cite{Titum1,CARPENTIER2015779}. The other part \(\rho^{(n)}(t)\) represents the combined effect of all propagators \(U_\Phi^{(m<n)}\). 
	
	The transported charge $Q^{(n)}$ can be separated into  two contributions: \(Q^{(n),\mathrm{diag}}\), which is diagonal in the respective Floquet states, and the off-diagonal \(Q^{(n),\mathrm{off-diag}}\). Persistent oscillations in the off-diagonal contribution prevent quantization of the transported charge \(Q^{(n)}\) over a single period. From Ref.~\cite{Titum1} we know that, without temporal disorder, one should consider the cumulative averaged pumped charge for which \(\frac{1}{N}\sum_{n=1}^NQ^{\mathrm{diag}}(N)=\mathrm{const.}\) while the off-diagonal contribution vanishes as \(\frac{1}{N}\sum_{n=1}^NQ^{\mathrm{off-diag}}(N)\sim N^{-\mu}\) with  \(\mu\leq1\) for large $N$.
	For an AFAI the constant cumulative averaged diagonal contribution to the transported charge is indeed quantized.
	
	For the present study, with temporal disorder, we also consider the cumulative charge average \(\frac{1}{N}\sum_{n=1}^N Q^{(n)}\). Still, the  oscillations of the off-diagonal part do not contribute in the long-time limit, as we have verified with numerical simulations not shown here. Note that only the cumulative time-averaged charge can be expected to show strict quantization, while the individual contributions $Q^{(n)}\) from each period may fluctuate around the quantized value~\cite{Titum1}.

	To analyze the quantity
	\begin{equation}
		\langle Q\rangle_N = \frac{1}{N}\sum_{n=1}^N Q^{(n),\mathrm{diag}}\,,
		\label{eq:ChargeCum}
	\end{equation}
	for the DFTI
	we again take a sequence \((\mathcal{J}^{(n)})\) from the distribution~\eqref{eq:JDist}.  As we have already seen that the value of $\mathcal{J}_0$ is not relevant for the transition to diffusive transport, we now fix \(\mathcal{J}_0=\pi/2\). The lattice geometry is chosen such that \(L_y=2L_x\), and we will vary the system size to check the finite-size effects. Here, the \(Q^{(n)}\) are averaged over $10^3$ disorder realizations as well as over all lines \(x=x_0\) (effectively corresponding to \(10^3 \times L_x\) disorder realizations). 
	\begin{figure}
		\centering
		\includegraphics{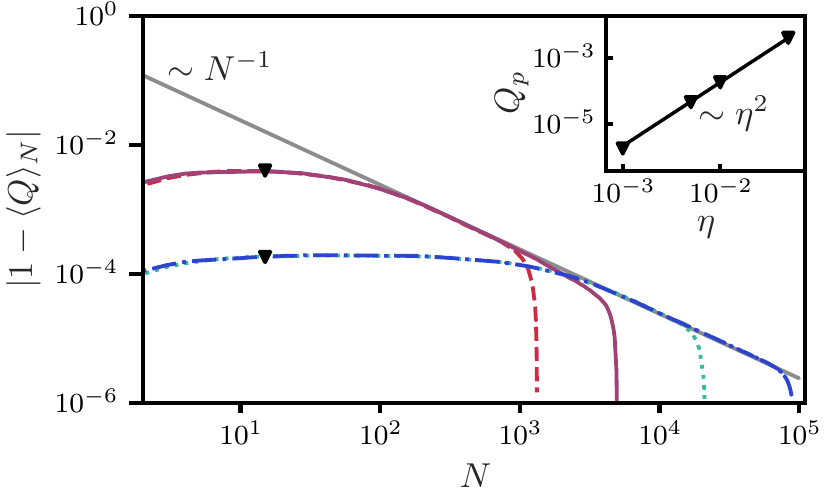}
		\caption{\label{fig:ChargeNum} Deviation of the cumulative averaged pumped charge from unity, \(|1-\langle Q \rangle_N|\), as a function of stroboscopic time \(N\) at various disorder strengths \(\eta\). Parameter sets used are: \((L_x,\eta)=(10,0.05)\) (red, dashed), \((L_x,\eta)=(20,0.05)\) (violet, solid), \((L_x,\eta)=(10,0.01)\) (green, dotted) and \((L_x,\eta)=(20,0.01)\) (blue, dot-dashed). The gray  line indicates a scaling behavior \(|1-\langle Q \rangle_N|\sim N^{-1}\). Inset: Plateau value \(Q_p\) as a function of disorder strength \(\eta\). Here, additional \(\eta\) values were included compared to the main plot.}
	\end{figure}
	
	We now investigate the deviation \(| 1- \langle Q \rangle_N|\) of the transported charge from the quantized value $\langle Q\rangle_N  = 1$. By analyzing the long-time behavior of \(| 1- \langle Q \rangle_N|\) in terms of scaling laws we can argue for the quantization of \(\langle Q\rangle_N\)  in the limit  \(L_x,N\to\infty\). 
	
	As seen in Fig.~\ref{fig:ChargeNum}, the charge $\langle Q\rangle_N$, equivalently the deviation \(|1-\langle Q \rangle_N|\), has a plateau over many periods $N$ (note the logarithmic axes in the figure). 
The plateau value \(Q_p\) given in the inset of Fig.~\ref{fig:ChargeNum} can be obtained as an average over the almost constant plateau. We observe that the plateau value, although being very close to $\langle Q\rangle_N=1$, is not strictly quantized.

Non-quantized plateaus for the pumped charge have been observed in Ref.~\cite{timms2021quantized} for a certain type of ``one-dimensional'' disorder. 
Since the ``one-dimensional'' disorder of Ref.~\cite{timms2021quantized} preserves translation symmetry in one direction, and thus does not lead to Anderson localization even without temporal fluctuations, it cannot be expected to result in any quantization. 

 In this situation, the question arises whether the plateaus observed here, for a fully ``two-dimensional'' disorder, result in strict quantization in the long-time limit. Note that, since only the cumulative charge $\langle Q\rangle_N$ may be quantized but not the pumped charge per period~\cite{Titum1}, a deviation from quantization in the transient regime does not contradict quantization in the long-time limit.

The numerical analysis of this question is complicated by finite-size effects which prevent direct access to the long-time limit. Instead, we can observe in our numerical data a characteristic scaling that emerges for times \(N\gg N_p\), where $N_p$ gives the plateau length and thus corresponds to the transition time between the transient regime and the long-time behavior. 
The scaling gives the deviations from quantization as \(| 1- \langle Q \rangle_N|\sim N^{-1}\).  As a guide to the eye, this scaling is included as the gray line in Fig.~\ref{fig:ChargeNum}.
Note how the curves follow this line over several orders of magnitude. The scaling $\sim N^{-1}$ is observed also in numerical data for parameters different from those used in Fig.~\ref{fig:ChargeNum}, which are not included here.
It appears that the scaling occurs as long as the disorder is not so strong as to destroy the topological phase.

 The limitation to finite system sizes eventually results in the breakdown of  quantization, and thus of the scaling \(| 1- \langle Q \rangle_N|\sim N^{-1}\). In Fig.~\ref{fig:ChargeNum}, the breakdown is visible as the apparent singularities that occur in our logarithmic plot when $1- \langle Q \rangle_N$ passes through zero. Data beyond this point have been omitted because they are entirely attributed to finite size effects and therefore have no relevance for quantization prior to this point.

The breakdown of quantization occurs as soon as the spread \(\sigma(N)\) of bulk states becomes comparable to the {height of the cylinder (recall that $L_y=2L_x$, cf. Fig.~\ref{fig:BulkSpread}), i.e., as soon as the corrections \(\sim e^{-L_x/\sigma(N)}\) are no longer negligible. From the estimate  \(\sigma(N)\simeq L_x\) and the scaling of the spread \(\sim\sqrt{N}\) we find \(N\sim L_x^2\), i.e., a doubling of the system size increases the accessible time scales by a factor of four. This is seen by comparison of the curves for $L_x = 10,20$ in Fig.~\ref{fig:ChargeNum}. 
Note that for the curve with \((L_x,\eta)=(20,0.01)\) the apparent singularity is already pushed towards the right end of the plot.
Could we perform our numerical computations for substantially larger system sizes, the singularities would no longer be visible in the plot. That shows that they are strictly finite-size effects.

Although we can neglect the finite-size effects only in the limit \(L_x\to\infty\), the given data indicate that with increasing $L_x$ the scaling $| 1- \langle Q \rangle_N|\sim N^{-1}$ continues for to ever longer, ultimately infinite, times.
 We can further observe that the long-time behavior of \(|1-\langle Q\rangle_N|\) for \(N\gg N_p\) is not only characterized by a common scaling \(\sim N^{-1}\), but that also the prefactor for different disorder strengths is the same. Therefore,  if the  disorder is not too strong, we can  expect that all deviations \(|1-\langle Q\rangle_N|\) from quantization of $\langle Q\rangle_N$ vanish as \(c N^{-1}\), with a unique constant $c$, in an infinite system ($L_x \to \infty$). If this is true it means that, apart from finite-size effects, the transported charge per period is exactly quantized in the long-time limit $N \to \infty$.
	
We can support this argument with a phenomenological relation between the deviations from quantization, the plateau value $Q_p$ and the transition time $N_p$ to diffusive transport, which reads
	\begin{equation}
		|1-\langle Q \rangle_N| = \frac{Q_p}{[1+(N/N_p)^\nu]^{1/\nu}}\,.
		\label{eq:PhenDev}
	\end{equation}	
	This functional relation describes a crossover from the constant plateau value \(Q_p\) for \(N\ll N_p\) to a value that scales as \(N^{-1}\) for \(N\gg N_p\).
	The parameter \(\nu\) controls the behavior in the crossover region. Equation~\eqref{eq:PhenDev} gives the constant $c$ of the previous scaling relation as \(c = Q_pN_p\), which is consistent with the inverse scaling of \(Q_p\) versus \(N_p\) observed in the numerical data. 
	
The inset of Fig.~\ref{fig:ChargeNum} shows that \(Q_p\sim \eta^{2}\), hence \(N_p = c/Q_p \sim \eta^{-2}\). In other words, the plateau length (i.e., the transition time) \(N_p\) for the charge transport scales exactly like the transition time \(N_c\) for reaching diffusive bulk transport. Note that the behavior \(Q_p\to 0\) and \(N_p\to \infty\) for \(\eta\to 0\), in the limit of vanishing temporal disorder, describes the persistent quantization known for the AFAI~\cite{Titum1}. 
	
	To conclude this section: In DFTIs with (for the moment: uncorrelated) temporal disorder numerical evidence for the scaling of the deviations from quantization strongly indicates that quantization of charge transport emerges in the long-time limit even in the absence of Anderson localization, i.e., for diffusive bulk transport.
	
	\section{Correlated temporal disorder}
	\begin{figure}
		\centering
		\includegraphics{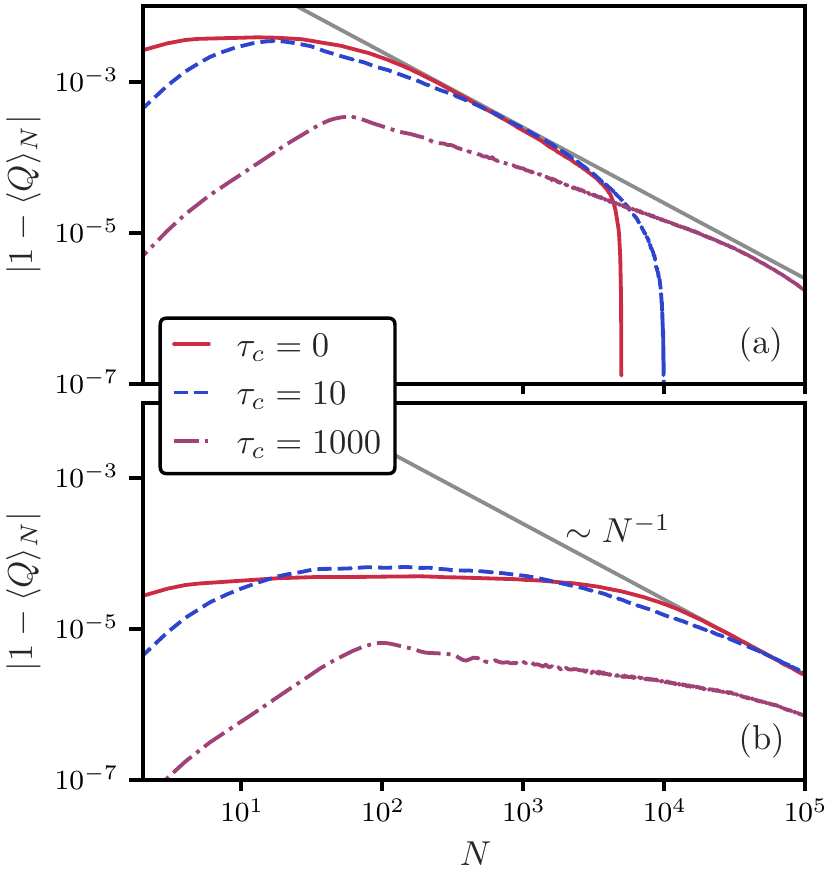}
		\caption{\label{fig:ChargeCorr} Deviation of the cumulative averaged pumped charge from unity, \(|1-\langle Q \rangle_N|\), as a function of stroboscopic time \(N\) for different correlation lengths \(\tau_c\). The disorder strength is \(\eta=0.05\) and \(\eta=0.005\) in (a) and (b), respectively. In both cases \(L_x=10\). The gray line indicates a scaling \(|1-\langle Q\rangle_N|\sim N^{-1}\).}
	\end{figure}
	In this section, we discuss the effect of temporally correlated disorder. We use disorder described  by the Ornstein-Uhlenbeck correlation function \cite{GILLESPIE1992111}
	\begin{equation}
		\mathrm{Corr}[\mathcal{J}^{(i)},\mathcal{J}^{(j)}]=e^{-|i-j|/\tau_c} \;.
		\label{eq:Corr}
	\end{equation}
	The correlation \(\mathrm{Corr}[X,Y]=\mathds{E}[(X-\mu_X)(Y-\mu_Y)]/\sigma_X\sigma_Y\) is defined as usual for two random variables \(X\) and \(Y\) with expectation value \(\mu\) and standard deviation \(\sigma\). The parameter \(\tau_c\) denotes the correlation time.
	
	For \(\tau_c\to\infty\), with perfect correlation over infinite times, we have \(\mathcal{J}^{(i)}=\mathcal{J}_0\) for all \(i\). This corresponds to vanishing temporal disorder.  For \(\tau_c\to 0\), we recover the uncorrelated case. Through variation of the correlation time \(\tau_c\) we can interpolate between uncorrelated temporal disorder, correlated disorder and a system without temporal disorder.

Figure~\ref{fig:ChargeCorr} compares the deviation of the cumulative averaged pumped charge from the quantized value $\langle Q \rangle_N = 1$ for correlated and uncorrelated temporal disorder.
In contrast to the uncorrelated disorder, with correlations a plateau can hardly be identified.
Nevertheless, for short correlation times ($\tau_c = 10$), the scaling \(|1-\langle Q\rangle_N | \sim N^{-1}\) emerges around the same as for the uncorrelated temporal disorder.
For very large correlation times ($\tau_c = 1000$) the transient regime is no longer comparable to the uncorrelated case \(\tau_c=0\) in the transient regime, but still the scaling \(|1-\langle Q\rangle | \sim N^{-1}\) seems to be obtained for very large times $N \simeq 10^4 \dots 10^5$.
With these data, we can safely assume that, as a consequence of this scaling,  the quantization of charge transport persists even for correlated temporal disorder in the long-time limit.
	
Figure~\ref{fig:ChargeCorr}~(b) shows that almost the same qualitative differences can be observed for weak disorder [\(\eta=0.005\) compared to \(\eta=0.05\) used in Fig.~\ref{fig:ChargeCorr}~(a)]. Here, the plateau is much broader and the crossover to the scaling \(|1-\langle Q\rangle_N | \sim N^{-1}\) takes place later.
 For very large correlation times (\(\tau_c=1000\)), the scaling might emerge for even larger times $N > 10^5$ which lie beyond the times accessible by our numerics. Much larger system sizes would be required to deal with the finite-size effects for such extreme times. 
	
	\section{\label{sec:Conclusion}Conclusions}
	To summarize, we have performed numerical simulations that give a clear indication that quantization of charge transport  in disordered Floquet topological insulators does not require Anderson localization. We arrive at this conclusion via an analysis of the deviation of the cumulative averaged pumped charge from unity, for which we find the scaling behavior \(|1-\langle Q \rangle_N| \sim N^{-1}\) with the number of periods.
	The observation of such a specific scaling (even with the same prefactor in different situations) allows us to extrapolate beyond times directly accessible in the numerics, which is necessarily restricted by finite-size effects.
In this way, the scaling implies that, for a system in the thermodynamic limit \(L_x\to\infty\), deviations from quantization vanish in the long-time limit \(N\to\infty\), i.e., the pumped charge is quantized.
 For uncorrelated temporal disorder we can even suggest a phenomenological relation describing the crossover of \(|1-\langle Q \rangle_N|\) from a plateau value to the scaling $\sim N^{-1}$ in a semi-quantitative fashion. For correlated temporal disorder the behavior for short correlation times is qualitatively similar to that for uncorrelated disorder, but becomes less distinct for large correlation times. Here, further investigation is required. However, in all situations the scaling \(|1-\langle Q \rangle_N| \sim N^{-1}\) emerges whenever we first observe a non-quantized plateau. 
In other words, a non-quantized plateau value of $\langle Q \rangle_N$ is associated with the transient regime, not the long-time limit.

The best scenario for the long-time limit compatible with the present numerical data is strict quantization of the cumulative averaged charge transport. 
Apparently, this is possible even without Anderson localization, which is destroyed by the temporal fluctuations of the disorder.
That quantization can survive the destruction of Anderson localization, which should be counterintuitive at first, is not entirely implausible: Through the temporal fluctuations the bulk becomes diffusive, with a scaling $\sigma \sim N^{1/2}$ for the spread of the bulk wave functions. This has to be compared with the spread $\sim N$ of topological edge states (recall the factor $N^{-1}$ in Eq.~\eqref{eq:ChargeCum} for the pumped charge). Only if the bulk became ballistic, with $\sigma \sim N$, quantization would be destroyed.

	In light of the present results further theoretical or experimental investigation of the \emph{disordered Floquet topological insulator}, which allows for temporal disorder, in contrast to the \emph{Anderson Floquet topological insulator}, which does not allow for temporal disorder, seems worthwhile. This could be of particular interest in the rapidly evolving  
		field  of quantum information applications, where symmetry-protected Floquet topological states \cite{PhysRevB.98.165421,zhang2022digital} play a major role. 
While extended numerical computations for increasingly larger system sizes might clarify the behavior for long correlation times that we can not resolve with the present numerics, it seems that an effective theoretical derivation of the scaling behavior instead of the simple phenomenological relation that we  provided here is equally important.
	
	\appendix
	
	\section{\label{appendix:missed}Missed-kicks disorder}
	\begin{figure}[b]
		\centering
		\includegraphics{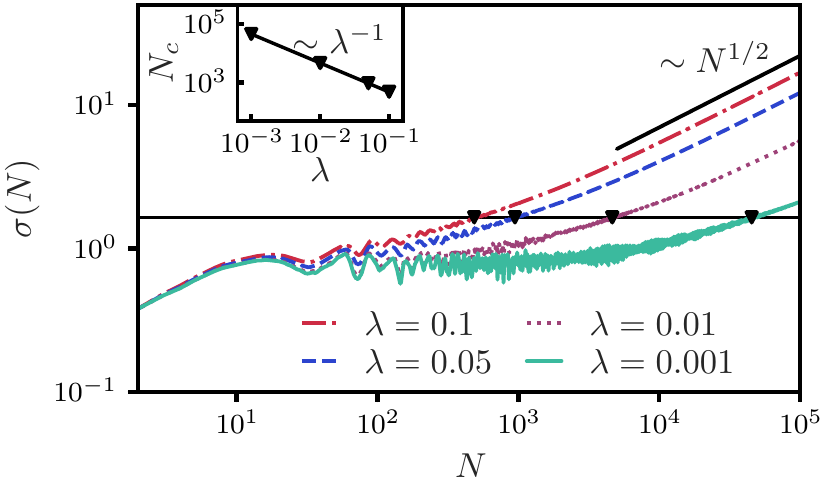}
		\caption{\label{fig:SigPoi} Spread \(\sigma(N)\) as a function of the stroboscopic time \(N\) at various parameters \(\lambda\) of the missed-kicks disorder. The horizontal black line marks \(\sigma(N_c)=3\) (arbitrary value) to estimate a critical time  \(N_c\) for the transition to diffusive bulk transport.}
	\end{figure}

	We now use a different kind of temporal disorder to demonstrate the wider validity of the results presented in the main text. 
	We consider the so-called missed-kicks disorder \cite{ravindranath2021dynamical,vcadevz2017dynamical}, where some of the $\delta$-phase-kicks are randomly omitted. The main mechanism for the loss of Anderson localization for the missed-kicks disorder is different from the mechanism for the frequency fluctuations used in the main text, see Ref.~\cite{ravindranath2021dynamical}. 
	
	For the missed-kicks disorder, the single-particle Hamiltonian of the \(n-\)th period reads \(H^{(n)}(t)=H_k+\delta(t-nT)g^{(n)}H_{\mathrm{stat-dis}}\) for \((k-1)\leq4[t/T-(n-1)]<k\) (\(k\in\{1,2,3,4\}\)), where \(H_k=\mathcal{J}\sum_{\bm{r}\in \Gamma}(c_{\mathbf{r}+\bm{\upgamma}_k}^\dagger c_{\mathbf{r}}+ c_{\mathbf{r}}^\dagger c_{\mathbf{r}+\bm{\upgamma}_k})\) and \(H_{\mathrm{stat-dis}}=\sum_{\mathbf{r}\in\Gamma}\phi_{\mathbf{r}}c_{\mathbf{r}}^\dagger c_{\mathbf{r}}\). The binary random variable \(g^{(n)}\) determines if a kick occurs or not. It is  defined by specifying the waiting times between two kicks: Given a sequence of waiting times \((\kappa^{(n)})\), if the \(l\)-th kick occurs in the \(m\)-th period (\(g^{(m)}=1\), \(l\leq m\)) the next kick occurs in the \(m+\kappa^{l}\)-th period. In other words, \(g^{(n)}=1\) holds if \(n\in (\kappa^{(1)},\kappa^{(1)}+\kappa^{(2)},\ldots)\) and \(g^{(n)}=0\) else. An appropriate choice for the distribution of waiting times is the Poisson distribution
	\begin{equation}
		P(\kappa^{(n)}=\kappa)=\frac{e^{-\lambda}\lambda^{\kappa-1}}{(\kappa-1)!}\,.
	\end{equation}
	If the disorder parameter \(\lambda<1\), missed kicks are rare.
	
		\begin{figure}[t]
		\centering
		\includegraphics{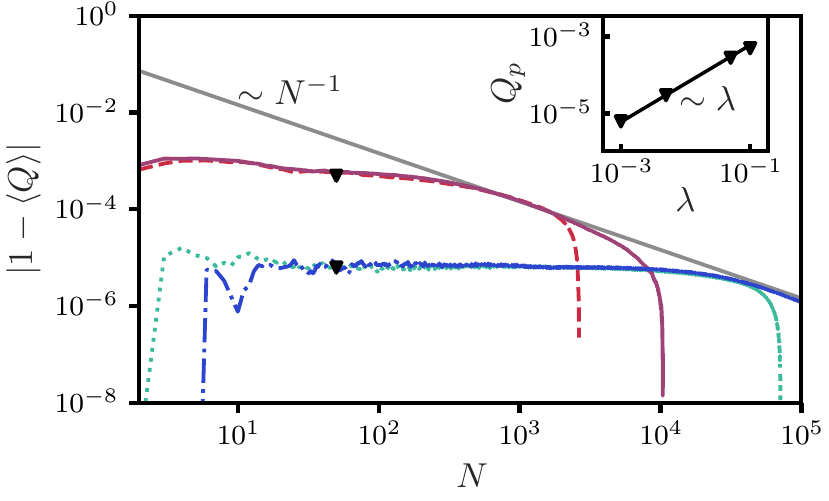}
		\caption{\label{fig:ChargePoiNum} Deviation  \(|1-\langle Q \rangle_N|\) of the cumulative averaged pumped charge from $\langle Q \rangle_N=1$ for the missed-kicks disorder, as a function of stroboscopic time \(N\) for various system sizes $L_x$ and disorder parameters \(\lambda\).  The parameters are: \((L_x,\lambda)=(10,0.1)\) (red, dashed), \((L_x,\lambda)=(20,0.1)\) (violet, solid), \((L_x,\lambda)=(10,0.001)\) (green, dotted) and \((L_x,\lambda)=(20,0.001)\) (blue, dot-dashed).
			The gray line indicates the scaling behavior \(\sim N^{-1}\). Inset: Plateau value \(Q_p\) of $\langle Q \rangle_N$ as a function of disorder strength \(\lambda\). Here, additional \(\lambda\) values are used compared to the main plot.
		}
	\end{figure}

	To monitor the dynamics of the system, we consider again the propagators \(U^{(n)}(T)=S_\phi^{(n)}U_4\ldots U_1\), where \(S_\phi^{(n)}\) equals \(S_\phi=e^{-iH_{\mathrm{stat-dis}}}\) if a kick occurs and \(\mathds{1}\) else.
	
	The methods and system geometries used to analyze the transport are the same as in Secs.~III and~IV. We now use fixed coupling constants \(\mathcal{J}=0.95\times \pi/2\). Note that for perfect coupling \(\mathcal{J}=\pi/2\) phase disorder has no effect. 
	
	We start with a brief discussion of bulk transport in a \(L_x \times L_y=80\times80\) geometry. Figure~\ref{fig:SigPoi} shows the same qualitative behavior of the spread $\sigma(N)$ as Fig.~\ref{fig:BulkSpread}~(b): For finite temporal disorder, the bulk transport undergoes a transition from quasi-localized to diffusive transport at a certain time \(N_c\). To analyze the dependence of $N_c$ on the disorder parameter \(\lambda\), we read off $N_c$ at an arbitrary value \(\sigma(N_c)=3\) where all curves show approximate diffusive behavior. The dependence of \(N_c\) on \(\lambda\) is given by the scaling \(N_c\sim \lambda^{-1}\), see Fig.~\ref{fig:SigPoi}.
	Interestingly, diffusive bulk transport sets in early if the probability for missed kicks is large. Large systems are needed to deal with this situation.
	
	The deviation \(|1-\langle Q\rangle_N|\) of the pumped charge from the quantized value $\langle Q \rangle_N=1$  is determined for a $L_x \times L_y$ geometry with \(L_y=2L_x\) and TBC.
	In Fig.~\ref{fig:ChargePoiNum} we observe the same signatures as for the uncorrelated frequency fluctuations (see Fig.~\ref{fig:ChargeNum}): A plateau up to a time \(N_p\) where a transition to the scaling \(|1-\langle Q\rangle_N| \sim N^{-1}\) takes place. As argued in the main text, the breakdown of the scaling vanishes in the limit \(L_x\to\infty\). The deviations of the pumped charge \(|1-\langle Q \rangle_N|\) agree with our phenomenological result Eq.~\eqref{eq:PhenDev} in the thermodynamic limit. The constant of the common scaling law in the long-time limit is again given by \(c=Q_pN_p\). The inset of Fig.~\ref{fig:ChargePoiNum} shows that \(Q_p\sim \lambda\), hence \(N_p=c/Q_p\sim \lambda^{-1}\) is equal to the scaling observed for reaching diffusive bulk transport (see inset of Fig.~\ref{fig:SigPoi}).

	The \(|1-\langle Q\rangle_N| \sim N^{-1}\) scaling implies that the transported charge becomes quantized in the long-time limit \(N\to\infty\). 
	Just as for the uncorrelated temporal disorder studied in the main text we again observe quantized charge transport without Anderson localization, now for correlated temporal disorder. It is reasonable to assume that this observation remains valid for other types of (un-)correlated temporal disorder.


%


\begin{thebibliography}{63}%
\makeatletter
\providecommand \@ifxundefined [1]{%
 \@ifx{#1\undefined}
}%
\providecommand \@ifnum [1]{%
 \ifnum #1\expandafter \@firstoftwo
 \else \expandafter \@secondoftwo
 \fi
}%
\providecommand \@ifx [1]{%
 \ifx #1\expandafter \@firstoftwo
 \else \expandafter \@secondoftwo
 \fi
}%
\providecommand \natexlab [1]{#1}%
\providecommand \enquote  [1]{``#1''}%
\providecommand \bibnamefont  [1]{#1}%
\providecommand \bibfnamefont [1]{#1}%
\providecommand \citenamefont [1]{#1}%
\providecommand \href@noop [0]{\@secondoftwo}%
\providecommand \href [0]{\begingroup \@sanitize@url \@href}%
\providecommand \@href[1]{\@@startlink{#1}\@@href}%
\providecommand \@@href[1]{\endgroup#1\@@endlink}%
\providecommand \@sanitize@url [0]{\catcode `\\12\catcode `\$12\catcode
  `\&12\catcode `\#12\catcode `\^12\catcode `\_12\catcode `\%12\relax}%
\providecommand \@@startlink[1]{}%
\providecommand \@@endlink[0]{}%
\providecommand \url  [0]{\begingroup\@sanitize@url \@url }%
\providecommand \@url [1]{\endgroup\@href {#1}{\urlprefix }}%
\providecommand \urlprefix  [0]{URL }%
\providecommand \Eprint [0]{\href }%
\providecommand \doibase [0]{https://doi.org/}%
\providecommand \selectlanguage [0]{\@gobble}%
\providecommand \bibinfo  [0]{\@secondoftwo}%
\providecommand \bibfield  [0]{\@secondoftwo}%
\providecommand \translation [1]{[#1]}%
\providecommand \BibitemOpen [0]{}%
\providecommand \bibitemStop [0]{}%
\providecommand \bibitemNoStop [0]{.\EOS\space}%
\providecommand \EOS [0]{\spacefactor3000\relax}%
\providecommand \BibitemShut  [1]{\csname bibitem#1\endcsname}%
\let\auto@bib@innerbib\@empty
\bibitem [{\citenamefont {von Klitzing}(1986)}]{RevModPhys.58.519}%
  \BibitemOpen
  \bibfield  {author} {\bibinfo {author} {\bibfnamefont {K.}~\bibnamefont {von
  Klitzing}},\ }\href {https://doi.org/10.1103/RevModPhys.58.519} {\bibfield
  {journal} {\bibinfo  {journal} {Rev. Mod. Phys.}\ }\textbf {\bibinfo {volume}
  {58}},\ \bibinfo {pages} {519} (\bibinfo {year} {1986})}\BibitemShut
  {NoStop}%
\bibitem [{\citenamefont {Thouless}(1983)}]{Thouless}%
  \BibitemOpen
  \bibfield  {author} {\bibinfo {author} {\bibfnamefont {D.~J.}\ \bibnamefont
  {Thouless}},\ }\href {https://doi.org/10.1103/PhysRevB.27.6083} {\bibfield
  {journal} {\bibinfo  {journal} {Phys. Rev. B}\ }\textbf {\bibinfo {volume}
  {27}},\ \bibinfo {pages} {6083} (\bibinfo {year} {1983})}\BibitemShut
  {NoStop}%
\bibitem [{\citenamefont {Thouless}\ \emph {et~al.}(1982)\citenamefont
  {Thouless}, \citenamefont {Kohmoto}, \citenamefont {Nightingale},\ and\
  \citenamefont {den Nijs}}]{TKNN}%
  \BibitemOpen
  \bibfield  {author} {\bibinfo {author} {\bibfnamefont {D.~J.}\ \bibnamefont
  {Thouless}}, \bibinfo {author} {\bibfnamefont {M.}~\bibnamefont {Kohmoto}},
  \bibinfo {author} {\bibfnamefont {M.~P.}\ \bibnamefont {Nightingale}},\ and\
  \bibinfo {author} {\bibfnamefont {M.}~\bibnamefont {den Nijs}},\ }\href
  {https://doi.org/10.1103/PhysRevLett.49.405} {\bibfield  {journal} {\bibinfo
  {journal} {Phys. Rev. Lett.}\ }\textbf {\bibinfo {volume} {49}},\ \bibinfo
  {pages} {405} (\bibinfo {year} {1982})}\BibitemShut {NoStop}%
\bibitem [{\citenamefont {Kitagawa}\ \emph {et~al.}(2010)\citenamefont
  {Kitagawa}, \citenamefont {Berg}, \citenamefont {Rudner},\ and\ \citenamefont
  {Demler}}]{PhysRevB.82.235114}%
  \BibitemOpen
  \bibfield  {author} {\bibinfo {author} {\bibfnamefont {T.}~\bibnamefont
  {Kitagawa}}, \bibinfo {author} {\bibfnamefont {E.}~\bibnamefont {Berg}},
  \bibinfo {author} {\bibfnamefont {M.}~\bibnamefont {Rudner}},\ and\ \bibinfo
  {author} {\bibfnamefont {E.}~\bibnamefont {Demler}},\ }\href
  {https://doi.org/10.1103/PhysRevB.82.235114} {\bibfield  {journal} {\bibinfo
  {journal} {Phys. Rev. B}\ }\textbf {\bibinfo {volume} {82}},\ \bibinfo
  {pages} {235114} (\bibinfo {year} {2010})}\BibitemShut {NoStop}%
\bibitem [{\citenamefont {Hasan}\ and\ \citenamefont
  {Kane}(2010)}]{HasanKane2010}%
  \BibitemOpen
  \bibfield  {author} {\bibinfo {author} {\bibfnamefont {M.~Z.}\ \bibnamefont
  {Hasan}}\ and\ \bibinfo {author} {\bibfnamefont {C.~L.}\ \bibnamefont
  {Kane}},\ }\href {https://doi.org/10.1103/RevModPhys.82.3045} {\bibfield
  {journal} {\bibinfo  {journal} {Rev. Mod. Phys.}\ }\textbf {\bibinfo {volume}
  {82}},\ \bibinfo {pages} {3045} (\bibinfo {year} {2010})}\BibitemShut
  {NoStop}%
\bibitem [{\citenamefont {Qi}\ and\ \citenamefont
  {Zhang}(2011)}]{RevModPhys.83.1057_11}%
  \BibitemOpen
  \bibfield  {author} {\bibinfo {author} {\bibfnamefont {X.-L.}\ \bibnamefont
  {Qi}}\ and\ \bibinfo {author} {\bibfnamefont {S.-C.}\ \bibnamefont {Zhang}},\
  }\href {https://doi.org/10.1103/RevModPhys.83.1057} {\bibfield  {journal}
  {\bibinfo  {journal} {Rev. Mod. Phys.}\ }\textbf {\bibinfo {volume} {83}},\
  \bibinfo {pages} {1057} (\bibinfo {year} {2011})}\BibitemShut {NoStop}%
\bibitem [{\citenamefont {Rudner}\ \emph {et~al.}(2013)\citenamefont {Rudner},
  \citenamefont {Lindner}, \citenamefont {Berg},\ and\ \citenamefont
  {Levin}}]{PhysRevX.3.031005}%
  \BibitemOpen
  \bibfield  {author} {\bibinfo {author} {\bibfnamefont {M.~S.}\ \bibnamefont
  {Rudner}}, \bibinfo {author} {\bibfnamefont {N.~H.}\ \bibnamefont {Lindner}},
  \bibinfo {author} {\bibfnamefont {E.}~\bibnamefont {Berg}},\ and\ \bibinfo
  {author} {\bibfnamefont {M.}~\bibnamefont {Levin}},\ }\href
  {https://doi.org/10.1103/PhysRevX.3.031005} {\bibfield  {journal} {\bibinfo
  {journal} {Phys. Rev. X}\ }\textbf {\bibinfo {volume} {3}},\ \bibinfo {pages}
  {031005} (\bibinfo {year} {2013})}\BibitemShut {NoStop}%
\bibitem [{\citenamefont {Nathan}\ and\ \citenamefont
  {Rudner}(2015)}]{nathan2015topological}%
  \BibitemOpen
  \bibfield  {author} {\bibinfo {author} {\bibfnamefont {F.}~\bibnamefont
  {Nathan}}\ and\ \bibinfo {author} {\bibfnamefont {M.~S.}\ \bibnamefont
  {Rudner}},\ }\href {https://doi.org/10.1088/1367-2630/17/12/125014}
  {\bibfield  {journal} {\bibinfo  {journal} {New J. Phys.}\ }\textbf {\bibinfo
  {volume} {17}},\ \bibinfo {pages} {125014} (\bibinfo {year}
  {2015})}\BibitemShut {NoStop}%
\bibitem [{\citenamefont {Chiu}\ \emph {et~al.}(2016)\citenamefont {Chiu},
  \citenamefont {Teo}, \citenamefont {Schnyder},\ and\ \citenamefont
  {Ryu}}]{RevModPhys.88.035005_16}%
  \BibitemOpen
  \bibfield  {author} {\bibinfo {author} {\bibfnamefont {C.-K.}\ \bibnamefont
  {Chiu}}, \bibinfo {author} {\bibfnamefont {J.~C.~Y.}\ \bibnamefont {Teo}},
  \bibinfo {author} {\bibfnamefont {A.~P.}\ \bibnamefont {Schnyder}},\ and\
  \bibinfo {author} {\bibfnamefont {S.}~\bibnamefont {Ryu}},\ }\href
  {https://doi.org/10.1103/RevModPhys.88.035005} {\bibfield  {journal}
  {\bibinfo  {journal} {Rev. Mod. Phys.}\ }\textbf {\bibinfo {volume} {88}},\
  \bibinfo {pages} {035005} (\bibinfo {year} {2016})}\BibitemShut {NoStop}%
\bibitem [{\citenamefont {Basov}\ \emph {et~al.}(2017)\citenamefont {Basov},
  \citenamefont {Averitt},\ and\ \citenamefont {Hsieh}}]{basov2017towards}%
  \BibitemOpen
  \bibfield  {author} {\bibinfo {author} {\bibfnamefont {D.}~\bibnamefont
  {Basov}}, \bibinfo {author} {\bibfnamefont {R.}~\bibnamefont {Averitt}},\
  and\ \bibinfo {author} {\bibfnamefont {D.}~\bibnamefont {Hsieh}},\ }\href
  {https://doi.org/10.1038/nmat5017} {\bibfield  {journal} {\bibinfo  {journal}
  {Nat. Mater.}\ }\textbf {\bibinfo {volume} {16}},\ \bibinfo {pages} {1077}
  (\bibinfo {year} {2017})}\BibitemShut {NoStop}%
\bibitem [{\citenamefont {Zhang}\ \emph {et~al.}(2017)\citenamefont {Zhang},
  \citenamefont {Hess}, \citenamefont {Kyprianidis}, \citenamefont {Becker},
  \citenamefont {Lee}, \citenamefont {Smith}, \citenamefont {Pagano},
  \citenamefont {Potirniche}, \citenamefont {Potter}, \citenamefont
  {Vishwanath} \emph {et~al.}}]{zhang2017observation}%
  \BibitemOpen
  \bibfield  {author} {\bibinfo {author} {\bibfnamefont {J.}~\bibnamefont
  {Zhang}}, \bibinfo {author} {\bibfnamefont {P.~W.}\ \bibnamefont {Hess}},
  \bibinfo {author} {\bibfnamefont {A.}~\bibnamefont {Kyprianidis}}, \bibinfo
  {author} {\bibfnamefont {P.}~\bibnamefont {Becker}}, \bibinfo {author}
  {\bibfnamefont {A.}~\bibnamefont {Lee}}, \bibinfo {author} {\bibfnamefont
  {J.}~\bibnamefont {Smith}}, \bibinfo {author} {\bibfnamefont
  {G.}~\bibnamefont {Pagano}}, \bibinfo {author} {\bibfnamefont {I.-D.}\
  \bibnamefont {Potirniche}}, \bibinfo {author} {\bibfnamefont {A.~C.}\
  \bibnamefont {Potter}}, \bibinfo {author} {\bibfnamefont {A.}~\bibnamefont
  {Vishwanath}}, \emph {et~al.},\ }\href {https://doi.org/10.1038/nature21413}
  {\bibfield  {journal} {\bibinfo  {journal} {Nature}\ }\textbf {\bibinfo
  {volume} {543}},\ \bibinfo {pages} {217} (\bibinfo {year}
  {2017})}\BibitemShut {NoStop}%
\bibitem [{\citenamefont {H{\"o}ckendorf}\ \emph {et~al.}(2018)\citenamefont
  {H{\"o}ckendorf}, \citenamefont {Alvermann},\ and\ \citenamefont
  {Fehske}}]{PhysRevB.97.045140}%
  \BibitemOpen
  \bibfield  {author} {\bibinfo {author} {\bibfnamefont {B.}~\bibnamefont
  {H{\"o}ckendorf}}, \bibinfo {author} {\bibfnamefont {A.}~\bibnamefont
  {Alvermann}},\ and\ \bibinfo {author} {\bibfnamefont {H.}~\bibnamefont
  {Fehske}},\ }\href {https://doi.org/10.1103/PhysRevB.97.045140} {\bibfield
  {journal} {\bibinfo  {journal} {Phys. Rev. B}\ }\textbf {\bibinfo {volume}
  {97}},\ \bibinfo {pages} {045140} (\bibinfo {year} {2018})}\BibitemShut
  {NoStop}%
\bibitem [{\citenamefont {Esin}\ \emph {et~al.}(2018)\citenamefont {Esin},
  \citenamefont {Rudner}, \citenamefont {Refael},\ and\ \citenamefont
  {Lindner}}]{PhysRevB.97.245401}%
  \BibitemOpen
  \bibfield  {author} {\bibinfo {author} {\bibfnamefont {I.}~\bibnamefont
  {Esin}}, \bibinfo {author} {\bibfnamefont {M.~S.}\ \bibnamefont {Rudner}},
  \bibinfo {author} {\bibfnamefont {G.}~\bibnamefont {Refael}},\ and\ \bibinfo
  {author} {\bibfnamefont {N.~H.}\ \bibnamefont {Lindner}},\ }\href
  {https://doi.org/10.1103/PhysRevB.97.245401} {\bibfield  {journal} {\bibinfo
  {journal} {Phys. Rev. B}\ }\textbf {\bibinfo {volume} {97}},\ \bibinfo
  {pages} {245401} (\bibinfo {year} {2018})}\BibitemShut {NoStop}%
\bibitem [{\citenamefont {Oka}\ and\ \citenamefont
  {Kitamura}(2019)}]{oka2019floquet}%
  \BibitemOpen
  \bibfield  {author} {\bibinfo {author} {\bibfnamefont {T.}~\bibnamefont
  {Oka}}\ and\ \bibinfo {author} {\bibfnamefont {S.}~\bibnamefont {Kitamura}},\
  }\href {https://doi.org/10.1146/annurev-conmatphys-031218-013423} {\bibfield
  {journal} {\bibinfo  {journal} {Annu. Rev. Condens. Matter Phys.}\ }\textbf
  {\bibinfo {volume} {10}},\ \bibinfo {pages} {387} (\bibinfo {year}
  {2019})}\BibitemShut {NoStop}%
\bibitem [{\citenamefont {H{\"o}ckendorf}\ \emph
  {et~al.}(2019{\natexlab{a}})\citenamefont {H{\"o}ckendorf}, \citenamefont
  {Alvermann},\ and\ \citenamefont {Fehske}}]{PhysRevLett.123.190403}%
  \BibitemOpen
  \bibfield  {author} {\bibinfo {author} {\bibfnamefont {B.}~\bibnamefont
  {H{\"o}ckendorf}}, \bibinfo {author} {\bibfnamefont {A.}~\bibnamefont
  {Alvermann}},\ and\ \bibinfo {author} {\bibfnamefont {H.}~\bibnamefont
  {Fehske}},\ }\href {https://doi.org/10.1103/PhysRevLett.123.190403}
  {\bibfield  {journal} {\bibinfo  {journal} {Phys. Rev. Lett.}\ }\textbf
  {\bibinfo {volume} {123}},\ \bibinfo {pages} {190403} (\bibinfo {year}
  {2019}{\natexlab{a}})}\BibitemShut {NoStop}%
\bibitem [{\citenamefont {Nathan}\ \emph {et~al.}(2019)\citenamefont {Nathan},
  \citenamefont {Abanin}, \citenamefont {Berg}, \citenamefont {Lindner},\ and\
  \citenamefont {Rudner}}]{PhysRevB.99.195133}%
  \BibitemOpen
  \bibfield  {author} {\bibinfo {author} {\bibfnamefont {F.}~\bibnamefont
  {Nathan}}, \bibinfo {author} {\bibfnamefont {D.}~\bibnamefont {Abanin}},
  \bibinfo {author} {\bibfnamefont {E.}~\bibnamefont {Berg}}, \bibinfo {author}
  {\bibfnamefont {N.~H.}\ \bibnamefont {Lindner}},\ and\ \bibinfo {author}
  {\bibfnamefont {M.~S.}\ \bibnamefont {Rudner}},\ }\href
  {https://doi.org/10.1103/PhysRevB.99.195133} {\bibfield  {journal} {\bibinfo
  {journal} {Phys. Rev. B}\ }\textbf {\bibinfo {volume} {99}},\ \bibinfo
  {pages} {195133} (\bibinfo {year} {2019})}\BibitemShut {NoStop}%
\bibitem [{\citenamefont {Fulga}\ \emph {et~al.}(2019)\citenamefont {Fulga},
  \citenamefont {Maksymenko}, \citenamefont {Rieder}, \citenamefont {Lindner},\
  and\ \citenamefont {Berg}}]{PhysRevB.99.235408}%
  \BibitemOpen
  \bibfield  {author} {\bibinfo {author} {\bibfnamefont {I.~C.}\ \bibnamefont
  {Fulga}}, \bibinfo {author} {\bibfnamefont {M.}~\bibnamefont {Maksymenko}},
  \bibinfo {author} {\bibfnamefont {M.~T.}\ \bibnamefont {Rieder}}, \bibinfo
  {author} {\bibfnamefont {N.~H.}\ \bibnamefont {Lindner}},\ and\ \bibinfo
  {author} {\bibfnamefont {E.}~\bibnamefont {Berg}},\ }\href
  {https://doi.org/10.1103/PhysRevB.99.235408} {\bibfield  {journal} {\bibinfo
  {journal} {Phys. Rev. B}\ }\textbf {\bibinfo {volume} {99}},\ \bibinfo
  {pages} {235408} (\bibinfo {year} {2019})}\BibitemShut {NoStop}%
\bibitem [{\citenamefont {H{\"o}ckendorf}\ \emph
  {et~al.}(2019{\natexlab{b}})\citenamefont {H{\"o}ckendorf}, \citenamefont
  {Alvermann},\ and\ \citenamefont {Fehske}}]{PhysRevB.99.245102}%
  \BibitemOpen
  \bibfield  {author} {\bibinfo {author} {\bibfnamefont {B.}~\bibnamefont
  {H{\"o}ckendorf}}, \bibinfo {author} {\bibfnamefont {A.}~\bibnamefont
  {Alvermann}},\ and\ \bibinfo {author} {\bibfnamefont {H.}~\bibnamefont
  {Fehske}},\ }\href {https://doi.org/10.1103/PhysRevB.99.245102} {\bibfield
  {journal} {\bibinfo  {journal} {Phys. Rev. B}\ }\textbf {\bibinfo {volume}
  {99}},\ \bibinfo {pages} {245102} (\bibinfo {year}
  {2019}{\natexlab{b}})}\BibitemShut {NoStop}%
\bibitem [{\citenamefont {Wintersperger}\ \emph {et~al.}(2020)\citenamefont
  {Wintersperger}, \citenamefont {Braun}, \citenamefont {{\"U}nal},
  \citenamefont {Eckardt}, \citenamefont {Liberto}, \citenamefont {Goldman},
  \citenamefont {Bloch},\ and\ \citenamefont
  {Aidelsburger}}]{wintersperger2020realization}%
  \BibitemOpen
  \bibfield  {author} {\bibinfo {author} {\bibfnamefont {K.}~\bibnamefont
  {Wintersperger}}, \bibinfo {author} {\bibfnamefont {C.}~\bibnamefont
  {Braun}}, \bibinfo {author} {\bibfnamefont {F.~N.}\ \bibnamefont {{\"U}nal}},
  \bibinfo {author} {\bibfnamefont {A.}~\bibnamefont {Eckardt}}, \bibinfo
  {author} {\bibfnamefont {M.~D.}\ \bibnamefont {Liberto}}, \bibinfo {author}
  {\bibfnamefont {N.}~\bibnamefont {Goldman}}, \bibinfo {author} {\bibfnamefont
  {I.}~\bibnamefont {Bloch}},\ and\ \bibinfo {author} {\bibfnamefont
  {M.}~\bibnamefont {Aidelsburger}},\ }\href
  {https://doi.org/10.1038/s41567-020-0949-y} {\bibfield  {journal} {\bibinfo
  {journal} {Nat. Phys.}\ }\textbf {\bibinfo {volume} {16}},\ \bibinfo {pages}
  {1058} (\bibinfo {year} {2020})}\BibitemShut {NoStop}%
\bibitem [{\citenamefont {Else}\ \emph {et~al.}(2020)\citenamefont {Else},
  \citenamefont {Ho},\ and\ \citenamefont {Dumitrescu}}]{PhysRevX.10.021032}%
  \BibitemOpen
  \bibfield  {author} {\bibinfo {author} {\bibfnamefont {D.~V.}\ \bibnamefont
  {Else}}, \bibinfo {author} {\bibfnamefont {W.~W.}\ \bibnamefont {Ho}},\ and\
  \bibinfo {author} {\bibfnamefont {P.~T.}\ \bibnamefont {Dumitrescu}},\ }\href
  {https://doi.org/10.1103/PhysRevX.10.021032} {\bibfield  {journal} {\bibinfo
  {journal} {Phys. Rev. X}\ }\textbf {\bibinfo {volume} {10}},\ \bibinfo
  {pages} {021032} (\bibinfo {year} {2020})}\BibitemShut {NoStop}%
\bibitem [{\citenamefont {Kundu}\ \emph {et~al.}(2020)\citenamefont {Kundu},
  \citenamefont {Rudner}, \citenamefont {Berg},\ and\ \citenamefont
  {Lindner}}]{PhysRevB.101.041403}%
  \BibitemOpen
  \bibfield  {author} {\bibinfo {author} {\bibfnamefont {A.}~\bibnamefont
  {Kundu}}, \bibinfo {author} {\bibfnamefont {M.}~\bibnamefont {Rudner}},
  \bibinfo {author} {\bibfnamefont {E.}~\bibnamefont {Berg}},\ and\ \bibinfo
  {author} {\bibfnamefont {N.~H.}\ \bibnamefont {Lindner}},\ }\href
  {https://doi.org/10.1103/PhysRevB.101.041403} {\bibfield  {journal} {\bibinfo
   {journal} {Phys. Rev. B}\ }\textbf {\bibinfo {volume} {101}},\ \bibinfo
  {pages} {041403} (\bibinfo {year} {2020})}\BibitemShut {NoStop}%
\bibitem [{\citenamefont {Fedorova}\ \emph {et~al.}(2020)\citenamefont
  {Fedorova}, \citenamefont {Qiu}, \citenamefont {Linden},\ and\ \citenamefont
  {Kroha}}]{Fedorova2020}%
  \BibitemOpen
  \bibfield  {author} {\bibinfo {author} {\bibfnamefont {Z.}~\bibnamefont
  {Fedorova}}, \bibinfo {author} {\bibfnamefont {H.}~\bibnamefont {Qiu}},
  \bibinfo {author} {\bibfnamefont {S.}~\bibnamefont {Linden}},\ and\ \bibinfo
  {author} {\bibfnamefont {J.}~\bibnamefont {Kroha}},\ }\href
  {https://doi.org/10.1038/s41467-020-17510-z} {\bibfield  {journal} {\bibinfo
  {journal} {Nat. Comm.}\ }\textbf {\bibinfo {volume} {11}},\ \bibinfo {pages}
  {3758} (\bibinfo {year} {2020})}\BibitemShut {NoStop}%
\bibitem [{\citenamefont {H{\"o}ckendorf}\ \emph {et~al.}(2020)\citenamefont
  {H{\"o}ckendorf}, \citenamefont {Alvermann},\ and\ \citenamefont
  {Fehske}}]{PhysRevResearch.2.023235}%
  \BibitemOpen
  \bibfield  {author} {\bibinfo {author} {\bibfnamefont {B.}~\bibnamefont
  {H{\"o}ckendorf}}, \bibinfo {author} {\bibfnamefont {A.}~\bibnamefont
  {Alvermann}},\ and\ \bibinfo {author} {\bibfnamefont {H.}~\bibnamefont
  {Fehske}},\ }\href {https://doi.org/10.1103/PhysRevResearch.2.023235}
  {\bibfield  {journal} {\bibinfo  {journal} {Phys. Rev. Res.}\ }\textbf
  {\bibinfo {volume} {2}},\ \bibinfo {pages} {023235} (\bibinfo {year}
  {2020})}\BibitemShut {NoStop}%
\bibitem [{\citenamefont {Bergholtz}\ \emph {et~al.}(2021)\citenamefont
  {Bergholtz}, \citenamefont {Budich},\ and\ \citenamefont
  {Kunst}}]{bergholtz2020exceptional}%
  \BibitemOpen
  \bibfield  {author} {\bibinfo {author} {\bibfnamefont {E.~J.}\ \bibnamefont
  {Bergholtz}}, \bibinfo {author} {\bibfnamefont {J.~C.}\ \bibnamefont
  {Budich}},\ and\ \bibinfo {author} {\bibfnamefont {F.~K.}\ \bibnamefont
  {Kunst}},\ }\href {https://doi.org/10.1103/RevModPhys.93.015005} {\bibfield
  {journal} {\bibinfo  {journal} {Rev. Mod. Phys}\ }\textbf {\bibinfo {volume}
  {93}},\ \bibinfo {pages} {15005} (\bibinfo {year} {2021})}\BibitemShut
  {NoStop}%
\bibitem [{\citenamefont {Dahlhaus}\ \emph {et~al.}(2015)\citenamefont
  {Dahlhaus}, \citenamefont {Fregoso},\ and\ \citenamefont
  {Moore}}]{dahlhaus2015magnetization}%
  \BibitemOpen
  \bibfield  {author} {\bibinfo {author} {\bibfnamefont {J.~P.}\ \bibnamefont
  {Dahlhaus}}, \bibinfo {author} {\bibfnamefont {B.~M.}\ \bibnamefont
  {Fregoso}},\ and\ \bibinfo {author} {\bibfnamefont {J.~E.}\ \bibnamefont
  {Moore}},\ }\href {https://doi.org/10.1103/PhysRevLett.114.246802} {\bibfield
   {journal} {\bibinfo  {journal} {Phys. Rev. Lett.}\ }\textbf {\bibinfo
  {volume} {114}},\ \bibinfo {pages} {246802} (\bibinfo {year}
  {2015})}\BibitemShut {NoStop}%
\bibitem [{\citenamefont {Kaladzhyan}\ \emph {et~al.}(2017)\citenamefont
  {Kaladzhyan}, \citenamefont {Simon},\ and\ \citenamefont
  {Trif}}]{kaladzhyan2017controlling}%
  \BibitemOpen
  \bibfield  {author} {\bibinfo {author} {\bibfnamefont {V.}~\bibnamefont
  {Kaladzhyan}}, \bibinfo {author} {\bibfnamefont {P.}~\bibnamefont {Simon}},\
  and\ \bibinfo {author} {\bibfnamefont {M.}~\bibnamefont {Trif}},\ }\href
  {https://doi.org/10.1103/PhysRevB.96.020507} {\bibfield  {journal} {\bibinfo
  {journal} {Phys. Rev. B}\ }\textbf {\bibinfo {volume} {96}},\ \bibinfo
  {pages} {020507} (\bibinfo {year} {2017})}\BibitemShut {NoStop}%
\bibitem [{\citenamefont {Dai}\ \emph {et~al.}(2018)\citenamefont {Dai},
  \citenamefont {Wang},\ and\ \citenamefont {Yi}}]{PhysRevA.98.013635}%
  \BibitemOpen
  \bibfield  {author} {\bibinfo {author} {\bibfnamefont {C.~M.}\ \bibnamefont
  {Dai}}, \bibinfo {author} {\bibfnamefont {W.}~\bibnamefont {Wang}},\ and\
  \bibinfo {author} {\bibfnamefont {X.~X.}\ \bibnamefont {Yi}},\ }\href
  {https://doi.org/10.1103/PhysRevA.98.013635} {\bibfield  {journal} {\bibinfo
  {journal} {Phys. Rev. A}\ }\textbf {\bibinfo {volume} {98}},\ \bibinfo
  {pages} {013635} (\bibinfo {year} {2018})}\BibitemShut {NoStop}%
\bibitem [{\citenamefont {Topp}\ \emph {et~al.}(2022)\citenamefont {Topp},
  \citenamefont {T{\"o}rm{\"a}}, \citenamefont {Kennes},\ and\ \citenamefont
  {Mitra}}]{topp2022orbital}%
  \BibitemOpen
  \bibfield  {author} {\bibinfo {author} {\bibfnamefont {G.~E.}\ \bibnamefont
  {Topp}}, \bibinfo {author} {\bibfnamefont {P.}~\bibnamefont {T{\"o}rm{\"a}}},
  \bibinfo {author} {\bibfnamefont {D.~M.}\ \bibnamefont {Kennes}},\ and\
  \bibinfo {author} {\bibfnamefont {A.}~\bibnamefont {Mitra}},\ }\href
  {https://doi.org/10.1103/PhysRevB.105.195426} {\bibfield  {journal} {\bibinfo
   {journal} {Phys. Rev. B}\ }\textbf {\bibinfo {volume} {105}},\ \bibinfo
  {pages} {195426} (\bibinfo {year} {2022})}\BibitemShut {NoStop}%
\bibitem [{\citenamefont {Rechtsman}\ \emph {et~al.}(2013)\citenamefont
  {Rechtsman}, \citenamefont {Zeuner}, \citenamefont {Plotnik}, \citenamefont
  {Lumer}, \citenamefont {Podolsky}, \citenamefont {Dreisow}, \citenamefont
  {Nolte}, \citenamefont {Segev},\ and\ \citenamefont
  {Szameit}}]{rechtsman2013photonic}%
  \BibitemOpen
  \bibfield  {author} {\bibinfo {author} {\bibfnamefont {M.~C.}\ \bibnamefont
  {Rechtsman}}, \bibinfo {author} {\bibfnamefont {J.~M.}\ \bibnamefont
  {Zeuner}}, \bibinfo {author} {\bibfnamefont {Y.}~\bibnamefont {Plotnik}},
  \bibinfo {author} {\bibfnamefont {Y.}~\bibnamefont {Lumer}}, \bibinfo
  {author} {\bibfnamefont {D.}~\bibnamefont {Podolsky}}, \bibinfo {author}
  {\bibfnamefont {F.}~\bibnamefont {Dreisow}}, \bibinfo {author} {\bibfnamefont
  {S.}~\bibnamefont {Nolte}}, \bibinfo {author} {\bibfnamefont
  {M.}~\bibnamefont {Segev}},\ and\ \bibinfo {author} {\bibfnamefont
  {A.}~\bibnamefont {Szameit}},\ }\href {https://doi.org/10.1038/nature12066}
  {\bibfield  {journal} {\bibinfo  {journal} {Nature}\ }\textbf {\bibinfo
  {volume} {496}},\ \bibinfo {pages} {196} (\bibinfo {year}
  {2013})}\BibitemShut {NoStop}%
\bibitem [{\citenamefont {Maczewsky}\ \emph {et~al.}(2017)\citenamefont
  {Maczewsky}, \citenamefont {Zeuner}, \citenamefont {Nolte},\ and\
  \citenamefont {Szameit}}]{maczewsky2017observation}%
  \BibitemOpen
  \bibfield  {author} {\bibinfo {author} {\bibfnamefont {L.~J.}\ \bibnamefont
  {Maczewsky}}, \bibinfo {author} {\bibfnamefont {J.~M.}\ \bibnamefont
  {Zeuner}}, \bibinfo {author} {\bibfnamefont {S.}~\bibnamefont {Nolte}},\ and\
  \bibinfo {author} {\bibfnamefont {A.}~\bibnamefont {Szameit}},\ }\href
  {https://doi.org/10.1038/ncomms13756} {\bibfield  {journal} {\bibinfo
  {journal} {Nat. Comm.}\ }\textbf {\bibinfo {volume} {8}},\ \bibinfo {pages}
  {1} (\bibinfo {year} {2017})}\BibitemShut {NoStop}%
\bibitem [{\citenamefont {Mukherjee}\ \emph {et~al.}(2017)\citenamefont
  {Mukherjee}, \citenamefont {Spracklen}, \citenamefont {Valiente},
  \citenamefont {Andersson}, \citenamefont {{\"O}hberg}, \citenamefont
  {Goldman},\ and\ \citenamefont {Thomson}}]{mukherjee2017experimental}%
  \BibitemOpen
  \bibfield  {author} {\bibinfo {author} {\bibfnamefont {S.}~\bibnamefont
  {Mukherjee}}, \bibinfo {author} {\bibfnamefont {A.}~\bibnamefont
  {Spracklen}}, \bibinfo {author} {\bibfnamefont {M.}~\bibnamefont {Valiente}},
  \bibinfo {author} {\bibfnamefont {E.}~\bibnamefont {Andersson}}, \bibinfo
  {author} {\bibfnamefont {P.}~\bibnamefont {{\"O}hberg}}, \bibinfo {author}
  {\bibfnamefont {N.}~\bibnamefont {Goldman}},\ and\ \bibinfo {author}
  {\bibfnamefont {R.~R.}\ \bibnamefont {Thomson}},\ }\href
  {https://doi.org/10.1038/ncomms13918} {\bibfield  {journal} {\bibinfo
  {journal} {Nature Communications}\ }\textbf {\bibinfo {volume} {8}},\
  \bibinfo {pages} {1} (\bibinfo {year} {2017})}\BibitemShut {NoStop}%
\bibitem [{\citenamefont {Ozawa}\ \emph {et~al.}(2019)\citenamefont {Ozawa},
  \citenamefont {Price}, \citenamefont {Amo}, \citenamefont {Goldman},
  \citenamefont {Hafezi}, \citenamefont {Lu}, \citenamefont {Rechtsman},
  \citenamefont {Schuster}, \citenamefont {Simon}, \citenamefont {Zilberberg},\
  and\ \citenamefont {Carusotto}}]{RevModPhys.91.015006_19}%
  \BibitemOpen
  \bibfield  {author} {\bibinfo {author} {\bibfnamefont {T.}~\bibnamefont
  {Ozawa}}, \bibinfo {author} {\bibfnamefont {H.~M.}\ \bibnamefont {Price}},
  \bibinfo {author} {\bibfnamefont {A.}~\bibnamefont {Amo}}, \bibinfo {author}
  {\bibfnamefont {N.}~\bibnamefont {Goldman}}, \bibinfo {author} {\bibfnamefont
  {M.}~\bibnamefont {Hafezi}}, \bibinfo {author} {\bibfnamefont
  {L.}~\bibnamefont {Lu}}, \bibinfo {author} {\bibfnamefont {M.~C.}\
  \bibnamefont {Rechtsman}}, \bibinfo {author} {\bibfnamefont {D.}~\bibnamefont
  {Schuster}}, \bibinfo {author} {\bibfnamefont {J.}~\bibnamefont {Simon}},
  \bibinfo {author} {\bibfnamefont {O.}~\bibnamefont {Zilberberg}},\ and\
  \bibinfo {author} {\bibfnamefont {I.}~\bibnamefont {Carusotto}},\ }\href
  {https://doi.org/10.1103/RevModPhys.91.015006} {\bibfield  {journal}
  {\bibinfo  {journal} {Rev. Mod. Phys.}\ }\textbf {\bibinfo {volume} {91}},\
  \bibinfo {pages} {015006} (\bibinfo {year} {2019})}\BibitemShut {NoStop}%
\bibitem [{\citenamefont {Maczewsky}\ \emph {et~al.}(2020)\citenamefont
  {Maczewsky}, \citenamefont {H{\"o}ckendorf}, \citenamefont {Kremer},
  \citenamefont {Biesenthal}, \citenamefont {Heinrich}, \citenamefont
  {Alvermann}, \citenamefont {Fehske},\ and\ \citenamefont
  {Szameit}}]{NatMat20}%
  \BibitemOpen
  \bibfield  {author} {\bibinfo {author} {\bibfnamefont {L.~J.}\ \bibnamefont
  {Maczewsky}}, \bibinfo {author} {\bibfnamefont {B.}~\bibnamefont
  {H{\"o}ckendorf}}, \bibinfo {author} {\bibfnamefont {M.}~\bibnamefont
  {Kremer}}, \bibinfo {author} {\bibfnamefont {T.}~\bibnamefont {Biesenthal}},
  \bibinfo {author} {\bibfnamefont {M.}~\bibnamefont {Heinrich}}, \bibinfo
  {author} {\bibfnamefont {A.}~\bibnamefont {Alvermann}}, \bibinfo {author}
  {\bibfnamefont {H.}~\bibnamefont {Fehske}},\ and\ \bibinfo {author}
  {\bibfnamefont {A.}~\bibnamefont {Szameit}},\ }\href
  {https://doi.org/10.1038/s41563-020-0641-8} {\bibfield  {journal} {\bibinfo
  {journal} {Nat. Mater.}\ }\textbf {\bibinfo {volume} {19}},\ \bibinfo {pages}
  {855} (\bibinfo {year} {2020})}\BibitemShut {NoStop}%
\bibitem [{\citenamefont {Fleury}\ \emph {et~al.}(2016)\citenamefont {Fleury},
  \citenamefont {Khanikaev},\ and\ \citenamefont {Alu}}]{fleury2016floquet}%
  \BibitemOpen
  \bibfield  {author} {\bibinfo {author} {\bibfnamefont {R.}~\bibnamefont
  {Fleury}}, \bibinfo {author} {\bibfnamefont {A.~B.}\ \bibnamefont
  {Khanikaev}},\ and\ \bibinfo {author} {\bibfnamefont {A.}~\bibnamefont
  {Alu}},\ }\href {https://doi.org/10.1038/ncomms11744} {\bibfield  {journal}
  {\bibinfo  {journal} {Nat. Comm.}\ }\textbf {\bibinfo {volume} {7}},\
  \bibinfo {pages} {1} (\bibinfo {year} {2016})}\BibitemShut {NoStop}%
\bibitem [{\citenamefont {Peng}\ \emph {et~al.}(2016)\citenamefont {Peng},
  \citenamefont {Qin}, \citenamefont {Zhao}, \citenamefont {Shen},
  \citenamefont {Xu}, \citenamefont {Bao}, \citenamefont {Jia},\ and\
  \citenamefont {Zhu}}]{peng2016experimental}%
  \BibitemOpen
  \bibfield  {author} {\bibinfo {author} {\bibfnamefont {Y.-G.}\ \bibnamefont
  {Peng}}, \bibinfo {author} {\bibfnamefont {C.-Z.}\ \bibnamefont {Qin}},
  \bibinfo {author} {\bibfnamefont {D.-G.}\ \bibnamefont {Zhao}}, \bibinfo
  {author} {\bibfnamefont {Y.-X.}\ \bibnamefont {Shen}}, \bibinfo {author}
  {\bibfnamefont {X.-Y.}\ \bibnamefont {Xu}}, \bibinfo {author} {\bibfnamefont
  {M.}~\bibnamefont {Bao}}, \bibinfo {author} {\bibfnamefont {H.}~\bibnamefont
  {Jia}},\ and\ \bibinfo {author} {\bibfnamefont {X.-F.}\ \bibnamefont {Zhu}},\
  }\href {https://doi.org/10.1038/ncomms13368} {\bibfield  {journal} {\bibinfo
  {journal} {Nat. Comm.}\ }\textbf {\bibinfo {volume} {7}},\ \bibinfo {pages}
  {1} (\bibinfo {year} {2016})}\BibitemShut {NoStop}%
\bibitem [{\citenamefont {Nagulu}\ \emph {et~al.}(2022)\citenamefont {Nagulu},
  \citenamefont {Ni}, \citenamefont {Kord}, \citenamefont {Tymchenko},
  \citenamefont {Garikapati}, \citenamefont {Al{\`u}},\ and\ \citenamefont
  {Krishnaswamy}}]{nagulu2022chip}%
  \BibitemOpen
  \bibfield  {author} {\bibinfo {author} {\bibfnamefont {A.}~\bibnamefont
  {Nagulu}}, \bibinfo {author} {\bibfnamefont {X.}~\bibnamefont {Ni}}, \bibinfo
  {author} {\bibfnamefont {A.}~\bibnamefont {Kord}}, \bibinfo {author}
  {\bibfnamefont {M.}~\bibnamefont {Tymchenko}}, \bibinfo {author}
  {\bibfnamefont {S.}~\bibnamefont {Garikapati}}, \bibinfo {author}
  {\bibfnamefont {A.}~\bibnamefont {Al{\`u}}},\ and\ \bibinfo {author}
  {\bibfnamefont {H.}~\bibnamefont {Krishnaswamy}},\ }\href
  {https://doi.org/10.1038/s41928-022-00751-9} {\bibfield  {journal} {\bibinfo
  {journal} {Nat. Electron.}\ }\textbf {\bibinfo {volume} {5}},\ \bibinfo
  {pages} {300} (\bibinfo {year} {2022})}\BibitemShut {NoStop}%
\bibitem [{\citenamefont {Kumar}\ \emph {et~al.}(2022)\citenamefont {Kumar},
  \citenamefont {Gupta},\ and\ \citenamefont {Singh}}]{kumar2022topological}%
  \BibitemOpen
  \bibfield  {author} {\bibinfo {author} {\bibfnamefont {A.}~\bibnamefont
  {Kumar}}, \bibinfo {author} {\bibfnamefont {M.}~\bibnamefont {Gupta}},\ and\
  \bibinfo {author} {\bibfnamefont {R.}~\bibnamefont {Singh}},\ }\href
  {https://doi.org/10.1038/s41928-022-00775-1} {\bibfield  {journal} {\bibinfo
  {journal} {Nat. Electron.}\ }\textbf {\bibinfo {volume} {5}},\ \bibinfo
  {pages} {261} (\bibinfo {year} {2022})}\BibitemShut {NoStop}%
\bibitem [{\citenamefont {Shih}\ and\ \citenamefont {Niu}(1994)}]{Shih}%
  \BibitemOpen
  \bibfield  {author} {\bibinfo {author} {\bibfnamefont {W.-K.}\ \bibnamefont
  {Shih}}\ and\ \bibinfo {author} {\bibfnamefont {Q.}~\bibnamefont {Niu}},\
  }\href {https://doi.org/10.1103/PhysRevB.50.11902} {\bibfield  {journal}
  {\bibinfo  {journal} {Phys. Rev. B}\ }\textbf {\bibinfo {volume} {50}},\
  \bibinfo {pages} {11902} (\bibinfo {year} {1994})}\BibitemShut {NoStop}%
\bibitem [{\citenamefont {Dehghani}\ and\ \citenamefont
  {Mitra}(2015)}]{PhysRevB.92.165111}%
  \BibitemOpen
  \bibfield  {author} {\bibinfo {author} {\bibfnamefont {H.}~\bibnamefont
  {Dehghani}}\ and\ \bibinfo {author} {\bibfnamefont {A.}~\bibnamefont
  {Mitra}},\ }\href {https://doi.org/10.1103/PhysRevB.92.165111} {\bibfield
  {journal} {\bibinfo  {journal} {Phys. Rev. B}\ }\textbf {\bibinfo {volume}
  {92}},\ \bibinfo {pages} {165111} (\bibinfo {year} {2015})}\BibitemShut
  {NoStop}%
\bibitem [{\citenamefont {Privitera}\ \emph {et~al.}(2018)\citenamefont
  {Privitera}, \citenamefont {Russomanno}, \citenamefont {Citro},\ and\
  \citenamefont {Santoro}}]{Privitera}%
  \BibitemOpen
  \bibfield  {author} {\bibinfo {author} {\bibfnamefont {L.}~\bibnamefont
  {Privitera}}, \bibinfo {author} {\bibfnamefont {A.}~\bibnamefont
  {Russomanno}}, \bibinfo {author} {\bibfnamefont {R.}~\bibnamefont {Citro}},\
  and\ \bibinfo {author} {\bibfnamefont {G.~E.}\ \bibnamefont {Santoro}},\
  }\href {https://doi.org/10.1103/PhysRevLett.120.106601} {\bibfield  {journal}
  {\bibinfo  {journal} {Phys. Rev. Lett.}\ }\textbf {\bibinfo {volume} {120}},\
  \bibinfo {pages} {106601} (\bibinfo {year} {2018})}\BibitemShut {NoStop}%
\bibitem [{\citenamefont {Titum}\ \emph {et~al.}(2016)\citenamefont {Titum},
  \citenamefont {Berg}, \citenamefont {Rudner}, \citenamefont {Refael},\ and\
  \citenamefont {Lindner}}]{Titum1}%
  \BibitemOpen
  \bibfield  {author} {\bibinfo {author} {\bibfnamefont {P.}~\bibnamefont
  {Titum}}, \bibinfo {author} {\bibfnamefont {E.}~\bibnamefont {Berg}},
  \bibinfo {author} {\bibfnamefont {M.~S.}\ \bibnamefont {Rudner}}, \bibinfo
  {author} {\bibfnamefont {G.}~\bibnamefont {Refael}},\ and\ \bibinfo {author}
  {\bibfnamefont {N.~H.}\ \bibnamefont {Lindner}},\ }\href
  {https://doi.org/10.1103/PhysRevX.6.021013} {\bibfield  {journal} {\bibinfo
  {journal} {Phys. Rev. X}\ }\textbf {\bibinfo {volume} {6}},\ \bibinfo {pages}
  {021013} (\bibinfo {year} {2016})}\BibitemShut {NoStop}%
\bibitem [{\citenamefont {Kim}\ \emph {et~al.}(2020)\citenamefont {Kim},
  \citenamefont {Bagrets}, \citenamefont {Micklitz},\ and\ \citenamefont
  {Altland}}]{PhysRevB.101.165401}%
  \BibitemOpen
  \bibfield  {author} {\bibinfo {author} {\bibfnamefont {K.~W.}\ \bibnamefont
  {Kim}}, \bibinfo {author} {\bibfnamefont {D.}~\bibnamefont {Bagrets}},
  \bibinfo {author} {\bibfnamefont {T.}~\bibnamefont {Micklitz}},\ and\
  \bibinfo {author} {\bibfnamefont {A.}~\bibnamefont {Altland}},\ }\href
  {https://doi.org/10.1103/PhysRevB.101.165401} {\bibfield  {journal} {\bibinfo
   {journal} {Phys. Rev. B}\ }\textbf {\bibinfo {volume} {101}},\ \bibinfo
  {pages} {165401} (\bibinfo {year} {2020})}\BibitemShut {NoStop}%
\bibitem [{\citenamefont {Liu}\ \emph {et~al.}(2020)\citenamefont {Liu},
  \citenamefont {Fulga},\ and\ \citenamefont
  {Asb\'oth}}]{PhysRevResearch.2.022048}%
  \BibitemOpen
  \bibfield  {author} {\bibinfo {author} {\bibfnamefont {H.}~\bibnamefont
  {Liu}}, \bibinfo {author} {\bibfnamefont {I.~C.}\ \bibnamefont {Fulga}},\
  and\ \bibinfo {author} {\bibfnamefont {J.~K.}\ \bibnamefont {Asb\'oth}},\
  }\href {https://doi.org/10.1103/PhysRevResearch.2.022048} {\bibfield
  {journal} {\bibinfo  {journal} {Phys. Rev. Res.}\ }\textbf {\bibinfo {volume}
  {2}},\ \bibinfo {pages} {022048} (\bibinfo {year} {2020})}\BibitemShut
  {NoStop}%
\bibitem [{\citenamefont {Timms}\ \emph {et~al.}(2021)\citenamefont {Timms},
  \citenamefont {Sieberer},\ and\ \citenamefont
  {Kolodrubetz}}]{timms2021quantized}%
  \BibitemOpen
  \bibfield  {author} {\bibinfo {author} {\bibfnamefont {C.~I.}\ \bibnamefont
  {Timms}}, \bibinfo {author} {\bibfnamefont {L.~M.}\ \bibnamefont
  {Sieberer}},\ and\ \bibinfo {author} {\bibfnamefont {M.~H.}\ \bibnamefont
  {Kolodrubetz}},\ }\href {https://doi.org/10.1103/PhysRevLett.127.270601}
  {\bibfield  {journal} {\bibinfo  {journal} {Phys. Rev. Lett.}\ }\textbf
  {\bibinfo {volume} {127}},\ \bibinfo {pages} {270601} (\bibinfo {year}
  {2021})}\BibitemShut {NoStop}%
\bibitem [{\citenamefont {Ammann}\ \emph {et~al.}(1998)\citenamefont {Ammann},
  \citenamefont {Gray}, \citenamefont {Shvarchuck},\ and\ \citenamefont
  {Christensen}}]{PhysRevLett.80.4111}%
  \BibitemOpen
  \bibfield  {author} {\bibinfo {author} {\bibfnamefont {H.}~\bibnamefont
  {Ammann}}, \bibinfo {author} {\bibfnamefont {R.}~\bibnamefont {Gray}},
  \bibinfo {author} {\bibfnamefont {I.}~\bibnamefont {Shvarchuck}},\ and\
  \bibinfo {author} {\bibfnamefont {N.}~\bibnamefont {Christensen}},\ }\href
  {https://doi.org/10.1103/PhysRevLett.80.4111} {\bibfield  {journal} {\bibinfo
   {journal} {Phys. Rev. Lett.}\ }\textbf {\bibinfo {volume} {80}},\ \bibinfo
  {pages} {4111} (\bibinfo {year} {1998})}\BibitemShut {NoStop}%
\bibitem [{\citenamefont {Klappauf}\ \emph {et~al.}(1998)\citenamefont
  {Klappauf}, \citenamefont {Oskay}, \citenamefont {Steck},\ and\ \citenamefont
  {Raizen}}]{PhysRevLett.81.1203}%
  \BibitemOpen
  \bibfield  {author} {\bibinfo {author} {\bibfnamefont {B.~G.}\ \bibnamefont
  {Klappauf}}, \bibinfo {author} {\bibfnamefont {W.~H.}\ \bibnamefont {Oskay}},
  \bibinfo {author} {\bibfnamefont {D.~A.}\ \bibnamefont {Steck}},\ and\
  \bibinfo {author} {\bibfnamefont {M.~G.}\ \bibnamefont {Raizen}},\ }\href
  {https://doi.org/10.1103/PhysRevLett.81.1203} {\bibfield  {journal} {\bibinfo
   {journal} {Phys. Rev. Lett.}\ }\textbf {\bibinfo {volume} {81}},\ \bibinfo
  {pages} {1203} (\bibinfo {year} {1998})}\BibitemShut {NoStop}%
\bibitem [{\citenamefont {Kendon}\ and\ \citenamefont
  {Tregenna}(2003)}]{PhysRevA.67.042315}%
  \BibitemOpen
  \bibfield  {author} {\bibinfo {author} {\bibfnamefont {V.}~\bibnamefont
  {Kendon}}\ and\ \bibinfo {author} {\bibfnamefont {B.}~\bibnamefont
  {Tregenna}},\ }\href {https://doi.org/10.1103/PhysRevA.67.042315} {\bibfield
  {journal} {\bibinfo  {journal} {Phys. Rev. A}\ }\textbf {\bibinfo {volume}
  {67}},\ \bibinfo {pages} {042315} (\bibinfo {year} {2003})}\BibitemShut
  {NoStop}%
\bibitem [{\citenamefont {Kendon}(2007)}]{kendon_2007}%
  \BibitemOpen
  \bibfield  {author} {\bibinfo {author} {\bibfnamefont {V.}~\bibnamefont
  {Kendon}},\ }\href {https://doi.org/10.1017/S0960129507006354} {\bibfield
  {journal} {\bibinfo  {journal} {Math. Struct. in Comp. Science}\ }\textbf
  {\bibinfo {volume} {17}},\ \bibinfo {pages} {1169} (\bibinfo {year}
  {2007})}\BibitemShut {NoStop}%
\bibitem [{\citenamefont {Yin}\ \emph {et~al.}(2008)\citenamefont {Yin},
  \citenamefont {Katsanos},\ and\ \citenamefont {Evangelou}}]{yin2008quantum}%
  \BibitemOpen
  \bibfield  {author} {\bibinfo {author} {\bibfnamefont {Y.}~\bibnamefont
  {Yin}}, \bibinfo {author} {\bibfnamefont {D.}~\bibnamefont {Katsanos}},\ and\
  \bibinfo {author} {\bibfnamefont {S.}~\bibnamefont {Evangelou}},\ }\href
  {https://doi.org/10.1103/PhysRevA.77.022302} {\bibfield  {journal} {\bibinfo
  {journal} {Phys. Rev. A}\ }\textbf {\bibinfo {volume} {77}},\ \bibinfo
  {pages} {022302} (\bibinfo {year} {2008})}\BibitemShut {NoStop}%
\bibitem [{\citenamefont {Schreiber}\ \emph {et~al.}(2011)\citenamefont
  {Schreiber}, \citenamefont {Cassemiro}, \citenamefont {Poto{\v{c}}ek},
  \citenamefont {G{\'a}bris}, \citenamefont {Jex},\ and\ \citenamefont
  {Silberhorn}}]{schreiber2011decoherence}%
  \BibitemOpen
  \bibfield  {author} {\bibinfo {author} {\bibfnamefont {A.}~\bibnamefont
  {Schreiber}}, \bibinfo {author} {\bibfnamefont {K.}~\bibnamefont
  {Cassemiro}}, \bibinfo {author} {\bibfnamefont {V.}~\bibnamefont
  {Poto{\v{c}}ek}}, \bibinfo {author} {\bibfnamefont {A.}~\bibnamefont
  {G{\'a}bris}}, \bibinfo {author} {\bibfnamefont {I.}~\bibnamefont {Jex}},\
  and\ \bibinfo {author} {\bibfnamefont {C.}~\bibnamefont {Silberhorn}},\
  }\href {https://doi.org/10.1103/PhysRevLett.106.180403} {\bibfield  {journal}
  {\bibinfo  {journal} {Phys. Rev. Lett.}\ }\textbf {\bibinfo {volume} {106}},\
  \bibinfo {pages} {180403} (\bibinfo {year} {2011})}\BibitemShut {NoStop}%
\bibitem [{\citenamefont {White}\ \emph {et~al.}(2014)\citenamefont {White},
  \citenamefont {Ruddell},\ and\ \citenamefont {Hoogerland}}]{white2014phase}%
  \BibitemOpen
  \bibfield  {author} {\bibinfo {author} {\bibfnamefont {D.}~\bibnamefont
  {White}}, \bibinfo {author} {\bibfnamefont {S.}~\bibnamefont {Ruddell}},\
  and\ \bibinfo {author} {\bibfnamefont {M.}~\bibnamefont {Hoogerland}},\
  }\href {https://doi.org/10.1088/1367-2630/16/11/113039} {\bibfield  {journal}
  {\bibinfo  {journal} {New J. Phys.}\ }\textbf {\bibinfo {volume} {16}},\
  \bibinfo {pages} {113039} (\bibinfo {year} {2014})}\BibitemShut {NoStop}%
\bibitem [{\citenamefont {{\v{C}}ade{\v{z}}}\ \emph {et~al.}(2017)\citenamefont
  {{\v{C}}ade{\v{z}}}, \citenamefont {Mondaini},\ and\ \citenamefont
  {Sacramento}}]{vcadevz2017dynamical}%
  \BibitemOpen
  \bibfield  {author} {\bibinfo {author} {\bibfnamefont {T.}~\bibnamefont
  {{\v{C}}ade{\v{z}}}}, \bibinfo {author} {\bibfnamefont {R.}~\bibnamefont
  {Mondaini}},\ and\ \bibinfo {author} {\bibfnamefont {P.~D.}\ \bibnamefont
  {Sacramento}},\ }\href {https://doi.org/10.1103/PhysRevB.96.144301}
  {\bibfield  {journal} {\bibinfo  {journal} {Phys. Rev. B}\ }\textbf {\bibinfo
  {volume} {96}},\ \bibinfo {pages} {144301} (\bibinfo {year}
  {2017})}\BibitemShut {NoStop}%
\bibitem [{\citenamefont {Sieberer}\ \emph {et~al.}(2018)\citenamefont
  {Sieberer}, \citenamefont {Rieder}, \citenamefont {Fischer},\ and\
  \citenamefont {Fulga}}]{PhysRevB.98.214301}%
  \BibitemOpen
  \bibfield  {author} {\bibinfo {author} {\bibfnamefont {L.~M.}\ \bibnamefont
  {Sieberer}}, \bibinfo {author} {\bibfnamefont {M.-T.}\ \bibnamefont
  {Rieder}}, \bibinfo {author} {\bibfnamefont {M.~H.}\ \bibnamefont
  {Fischer}},\ and\ \bibinfo {author} {\bibfnamefont {I.~C.}\ \bibnamefont
  {Fulga}},\ }\href {https://doi.org/10.1103/PhysRevB.98.214301} {\bibfield
  {journal} {\bibinfo  {journal} {Phys. Rev. B}\ }\textbf {\bibinfo {volume}
  {98}},\ \bibinfo {pages} {214301} (\bibinfo {year} {2018})}\BibitemShut
  {NoStop}%
\bibitem [{\citenamefont {Rieder}\ \emph {et~al.}(2018)\citenamefont {Rieder},
  \citenamefont {Sieberer}, \citenamefont {Fischer},\ and\ \citenamefont
  {Fulga}}]{PhysRevLett.120.216801}%
  \BibitemOpen
  \bibfield  {author} {\bibinfo {author} {\bibfnamefont {M.-T.}\ \bibnamefont
  {Rieder}}, \bibinfo {author} {\bibfnamefont {L.~M.}\ \bibnamefont
  {Sieberer}}, \bibinfo {author} {\bibfnamefont {M.~H.}\ \bibnamefont
  {Fischer}},\ and\ \bibinfo {author} {\bibfnamefont {I.~C.}\ \bibnamefont
  {Fulga}},\ }\href {https://doi.org/10.1103/PhysRevLett.120.216801} {\bibfield
   {journal} {\bibinfo  {journal} {Phys. Rev. Lett.}\ }\textbf {\bibinfo
  {volume} {120}},\ \bibinfo {pages} {216801} (\bibinfo {year}
  {2018})}\BibitemShut {NoStop}%
\bibitem [{\citenamefont {Ravindranath}\ and\ \citenamefont
  {Santhanam}(2021)}]{ravindranath2021dynamical}%
  \BibitemOpen
  \bibfield  {author} {\bibinfo {author} {\bibfnamefont {V.}~\bibnamefont
  {Ravindranath}}\ and\ \bibinfo {author} {\bibfnamefont {M.}~\bibnamefont
  {Santhanam}},\ }\href {https://doi.org/10.1103/PhysRevB.103.134303}
  {\bibfield  {journal} {\bibinfo  {journal} {Phys. Rev. B}\ }\textbf {\bibinfo
  {volume} {103}},\ \bibinfo {pages} {134303} (\bibinfo {year}
  {2021})}\BibitemShut {NoStop}%
\bibitem [{\citenamefont {Cao}\ \emph {et~al.}(2022)\citenamefont {Cao},
  \citenamefont {Sajjad}, \citenamefont {Mas}, \citenamefont {Simmons},
  \citenamefont {Tanlimco}, \citenamefont {Nolasco-Martinez}, \citenamefont
  {Shimasaki}, \citenamefont {Kondakci}, \citenamefont {Galitski},\ and\
  \citenamefont {Weld}}]{cao2021interaction}%
  \BibitemOpen
  \bibfield  {author} {\bibinfo {author} {\bibfnamefont {A.}~\bibnamefont
  {Cao}}, \bibinfo {author} {\bibfnamefont {R.}~\bibnamefont {Sajjad}},
  \bibinfo {author} {\bibfnamefont {H.}~\bibnamefont {Mas}}, \bibinfo {author}
  {\bibfnamefont {E.~Q.}\ \bibnamefont {Simmons}}, \bibinfo {author}
  {\bibfnamefont {J.~L.}\ \bibnamefont {Tanlimco}}, \bibinfo {author}
  {\bibfnamefont {E.}~\bibnamefont {Nolasco-Martinez}}, \bibinfo {author}
  {\bibfnamefont {T.}~\bibnamefont {Shimasaki}}, \bibinfo {author}
  {\bibfnamefont {H.}~\bibnamefont {Kondakci}}, \bibinfo {author}
  {\bibfnamefont {V.}~\bibnamefont {Galitski}},\ and\ \bibinfo {author}
  {\bibfnamefont {D.~M.}\ \bibnamefont {Weld}},\ }\href
  {https://doi.org/10.1038/s41567-022-01724-7} {\bibfield  {journal} {\bibinfo
  {journal} {Nat. Phys.}\ }\textbf {\bibinfo {volume} {18}},\ \bibinfo {pages}
  {1302} (\bibinfo {year} {2022})}\BibitemShut {NoStop}%
\bibitem [{\citenamefont {Hamza}\ \emph {et~al.}(2009)\citenamefont {Hamza},
  \citenamefont {Joye},\ and\ \citenamefont {Stolz}}]{hamza2009dynamical}%
  \BibitemOpen
  \bibfield  {author} {\bibinfo {author} {\bibfnamefont {E.}~\bibnamefont
  {Hamza}}, \bibinfo {author} {\bibfnamefont {A.}~\bibnamefont {Joye}},\ and\
  \bibinfo {author} {\bibfnamefont {G.}~\bibnamefont {Stolz}},\ }\href
  {https://doi.org/10.1007/s11040-009-9068-9} {\bibfield  {journal} {\bibinfo
  {journal} {Math. Phys. Anal. Geom.}\ }\textbf {\bibinfo {volume} {12}},\
  \bibinfo {pages} {381} (\bibinfo {year} {2009})}\BibitemShut {NoStop}%
\bibitem [{\citenamefont {Metzler}\ \emph {et~al.}(2014)\citenamefont
  {Metzler}, \citenamefont {Jeon}, \citenamefont {Cherstvy},\ and\
  \citenamefont {Barkai}}]{metzler2014anomalous}%
  \BibitemOpen
  \bibfield  {author} {\bibinfo {author} {\bibfnamefont {R.}~\bibnamefont
  {Metzler}}, \bibinfo {author} {\bibfnamefont {J.-H.}\ \bibnamefont {Jeon}},
  \bibinfo {author} {\bibfnamefont {A.~G.}\ \bibnamefont {Cherstvy}},\ and\
  \bibinfo {author} {\bibfnamefont {E.}~\bibnamefont {Barkai}},\ }\href
  {https://doi.org/10.1039/C4CP03465A} {\bibfield  {journal} {\bibinfo
  {journal} {Phys. Chem. Chem. Phys.}\ }\textbf {\bibinfo {volume} {16}},\
  \bibinfo {pages} {24128} (\bibinfo {year} {2014})}\BibitemShut {NoStop}%
\bibitem [{\citenamefont {Niu}\ and\ \citenamefont
  {Thouless}(1984)}]{Thouless2}%
  \BibitemOpen
  \bibfield  {author} {\bibinfo {author} {\bibfnamefont {Q.}~\bibnamefont
  {Niu}}\ and\ \bibinfo {author} {\bibfnamefont {D.}~\bibnamefont {Thouless}},\
  }\href {https://doi.org/10.1088/0305-4470/17/12/016} {\bibfield  {journal}
  {\bibinfo  {journal} {J. Phys. A}\ }\textbf {\bibinfo {volume} {17}},\
  \bibinfo {pages} {2453} (\bibinfo {year} {1984})}\BibitemShut {NoStop}%
\bibitem [{\citenamefont {Carpentier}\ \emph {et~al.}(2015)\citenamefont
  {Carpentier}, \citenamefont {Delplace}, \citenamefont {Fruchart},
  \citenamefont {Gaw{ę}dzki},\ and\ \citenamefont
  {Tauber}}]{CARPENTIER2015779}%
  \BibitemOpen
  \bibfield  {author} {\bibinfo {author} {\bibfnamefont {D.}~\bibnamefont
  {Carpentier}}, \bibinfo {author} {\bibfnamefont {P.}~\bibnamefont
  {Delplace}}, \bibinfo {author} {\bibfnamefont {M.}~\bibnamefont {Fruchart}},
  \bibinfo {author} {\bibfnamefont {K.}~\bibnamefont {Gaw{ę}dzki}},\ and\
  \bibinfo {author} {\bibfnamefont {C.}~\bibnamefont {Tauber}},\ }\href
  {https://doi.org/https://doi.org/10.1016/j.nuclphysb.2015.05.009} {\bibfield
  {journal} {\bibinfo  {journal} {Nucl. Phys. B}\ }\textbf {\bibinfo {volume}
  {896}},\ \bibinfo {pages} {779} (\bibinfo {year} {2015})}\BibitemShut
  {NoStop}%
\bibitem [{\citenamefont {Gillespie}(1992)}]{GILLESPIE1992111}%
  \BibitemOpen
  \bibfield  {author} {\bibinfo {author} {\bibfnamefont {D.~T.}\ \bibnamefont
  {Gillespie}},\ }in\ \href
  {https://doi.org/https://doi.org/10.1016/B978-0-08-091837-2.50008-9} {\emph
  {\bibinfo {booktitle} {Markov Processes}}},\ \bibinfo {editor} {edited by\
  \bibinfo {editor} {\bibfnamefont {D.~T.}\ \bibnamefont {Gillespie}}}\
  (\bibinfo  {publisher} {Academic Press},\ \bibinfo {address} {San Diego},\
  \bibinfo {year} {1992})\ pp.\ \bibinfo {pages} {111--219}\BibitemShut
  {NoStop}%
\bibitem [{\citenamefont {Bomantara}\ and\ \citenamefont
  {Gong}(2018)}]{PhysRevB.98.165421}%
  \BibitemOpen
  \bibfield  {author} {\bibinfo {author} {\bibfnamefont {R.~W.}\ \bibnamefont
  {Bomantara}}\ and\ \bibinfo {author} {\bibfnamefont {J.}~\bibnamefont
  {Gong}},\ }\href {https://doi.org/10.1103/PhysRevB.98.165421} {\bibfield
  {journal} {\bibinfo  {journal} {Phys. Rev. B}\ }\textbf {\bibinfo {volume}
  {98}},\ \bibinfo {pages} {165421} (\bibinfo {year} {2018})}\BibitemShut
  {NoStop}%
\bibitem [{\citenamefont {Zhang}\ \emph {et~al.}(2022)\citenamefont {Zhang},
  \citenamefont {Jiang}, \citenamefont {Deng}, \citenamefont {Wang},
  \citenamefont {Chen}, \citenamefont {Zhang}, \citenamefont {Ren},
  \citenamefont {Dong}, \citenamefont {Xu}, \citenamefont {Gao} \emph
  {et~al.}}]{zhang2022digital}%
  \BibitemOpen
  \bibfield  {author} {\bibinfo {author} {\bibfnamefont {X.}~\bibnamefont
  {Zhang}}, \bibinfo {author} {\bibfnamefont {W.}~\bibnamefont {Jiang}},
  \bibinfo {author} {\bibfnamefont {J.}~\bibnamefont {Deng}}, \bibinfo {author}
  {\bibfnamefont {K.}~\bibnamefont {Wang}}, \bibinfo {author} {\bibfnamefont
  {J.}~\bibnamefont {Chen}}, \bibinfo {author} {\bibfnamefont {P.}~\bibnamefont
  {Zhang}}, \bibinfo {author} {\bibfnamefont {W.}~\bibnamefont {Ren}}, \bibinfo
  {author} {\bibfnamefont {H.}~\bibnamefont {Dong}}, \bibinfo {author}
  {\bibfnamefont {S.}~\bibnamefont {Xu}}, \bibinfo {author} {\bibfnamefont
  {Y.}~\bibnamefont {Gao}}, \emph {et~al.},\ }\href
  {https://doi.org/10.1038/s41586-022-04854-3} {\bibfield  {journal} {\bibinfo
  {journal} {Nature}\ }\textbf {\bibinfo {volume} {607}},\ \bibinfo {pages}
  {468} (\bibinfo {year} {2022})}\BibitemShut {NoStop}%
\end{thebibliography}
\end{document}